\documentclass[aps,pra,showpacs,twocolumn,amsmath,amssymb,superscriptaddress,footinbib]{revtex4}

\usepackage[english]{babel}
\usepackage{latexsym}
\usepackage{graphicx}
\usepackage{subfigure}
\usepackage{epsfig}
\usepackage{amsfonts}
\usepackage{amssymb}
\usepackage{amsmath}
\usepackage{bbm}

\begin{document}

\title{Chaos in circuit QED: decoherence, localization, and nonclassicality}


\author{Jonas Larson}
\email{jolarson@physto.su.se} \affiliation{Department of Physics,
Stockholm University, AlbaNova University Center, Se-106 91 Stockholm,
Sweden}
\affiliation{Institut f\"ur Theoretische Physik, Universit\"at zu K\"oln, K\"oln, De-50937, Germany}
\author{Duncan H. J. O'Dell}
\affiliation{Department of Physics and Astronomy, McMaster University, 1280 Main St. W., Hamilton, ON, L8S 4M1, Canada}

\date{\today}

\begin{abstract}
We study the open system dynamics of a circuit QED model operating in the ultrastrong coupling regime. If the resonator is pumped periodically in time the underlying classical system is chaotic. Indeed, the periodically driven Jaynes-Cummings model in the Born-Oppenheimer approximation resembles a Duffing oscillator which in the classical limit is a well-known example of a chaotic system. Detection of the field quadrature of the output field acts as an effective position measurement of the oscillator. We address how such detection affects the quantum chaotic evolution in this bipartite system. We differentiate between single measurement realizations and ensembles of repeated measurements. In the former case a measurement/decoherence induced localization effect is encountered, while in the latter this localization is almost completely absent. This is in marked contrast to numerous earlier works discussing the quantum-classical correspondence in measured chaotic systems. This lack of a classical correspondence under relatively strong measurement induced decoherence is attributed to the inherent quantum nature of the qubit subsystem and in particular to the quantum correlations between the qubit and the field which persist despite the decoherence.
\end{abstract}

\maketitle

\section{Introduction}\label{intro}
While chaos is well defined in classical mechanics, the degree to which it can manifest itself in quantum mechanics remains a problem of fundamental interest \cite{haake}. For example, classically chaotic systems exhibit exponential sensitivity to initial conditions, something that is absent from quantum dynamics which is explicitly unitary. Moreover, the discrete spectrum of closed quantum systems implies quasi-periodicity which contradicts the irregular behaviour seen in classical chaotic systems. Theoretical work has indicated that the inclusion of decoherence, on the other hand, can cause a quantum system to become exponentially sensitive to initial state fluctuations~\cite{decochaos}. 

On the experimental front, the advent of the laser cooling of atoms has allowed the study of a number of quantum chaotic phenomena in simple and highly controllable settings. Of particular relevance to this paper we highlight the observation of dynamical localization of momentum in a kicked quantum rotor \cite{moore94} made from atoms in a frequency modulated optical lattice, where classical (chaotic) diffusion is suppressed by quantum interference, and chaos assisted tunneling \cite{steck01,hensinger01}, which was observed using atoms in an amplitude modulated lattice. In the latter system the tunneling is dynamical in nature in the sense that it occurs in phase space between islands of regular motion, allowing the system to pass through the classically insurmountable dynamical ``barriers'' provided by the boundaries between integrable and chaotic regions. The kicked quantum top has also been realized using atomic spins \cite{chaudhury09}: using continuous weak measurement techniques, the full quantum dynamics in phase space can be reconstructed as a function of time. Some of these experimental setups have also been used to investigate the effects of decoherence \cite{klappauf98,steck02}, and, as expected, it was found that in each case the system reverted to classical behaviour.

Driven by the promise of technological applications, continuing progress in the manufacture of man made and hybrid quantum systems also provides significant opportunities for exploring basic quantum dynamics under controllable conditions~\cite{milburn}. Within the realm of this progress the idea of {\it quantum simulators} has become reality~\cite{maciek,trapsim,cavitysim}. While quantum simulation of closed systems has been the main focus thus far, the extension to open systems has been considered~\cite{opensim}. In our view these new quantum systems naturally lend themselves to the study of chaotic quantum dynamics subject to decoherence.

In this work we explore quantum chaos in a {\it circuit quantum electrodynamics} (QED) setup comprised of a superconducting qubit resonantly coupled to a single mode of a high-$Q$ transmission line resonator~\cite{circuitQED}.  The presence of the two-level qubit means that this is a manifestly quantum system in the sense that it does not possess a simple classical limit. The parameters of the corresponding Hamiltonian can be externally controlled to a large extent, and even the strength of the decoherence can be manipulated by various means. In order to achieve chaotic dynamics we consider the case where the resonator is externally pumped by a drive oscillator with adjustable amplitude and phase. By changing either the qubit-field coupling~\cite{solano1}, the pump amplitude or the pump modulation frequency, the system can be varied between being regular or chaotic. In particular, in the {\it ultrastrong coupling regime}~\cite{solano1}, and at vigourous enough pumping, the system becomes chaotic in the sense that the semi-classical equations of motion corresponding to a classical electromagnetic field (but retaining the quantum nature of the qubit), become chaotic.  Furthermore, in the Born-Oppenheimer approximation (BOA) where the excited state of the qubit is adiabatically eliminated, we find that the driven circuit QED system mimics  a quantum {\it Duffing oscillator}~\cite{duffing}. The duffing oscillator is an example of a well-studied system whose classical limit is known to be chaotic.
 
 In the chaotic regime we shall see that an initially localized phase space distribution of the electromagnetic field rapidly broadens and the Wigner distribution builds up seemingly irregular sub-Planck structures characteristic of chaos. In comparison to how a localized phase space distribution evolves under the corresponding semi-classical equations of motion, the broadening in the fully quantum evolution is much more dramatic leading to a breakdown of the quantum-classical correspondence. One possible route to recover the classical dynamics is to try to localize the state by weakly measuring the position quadrature of the field. This induces decoherence and non-unitary time evolution~\cite{decomeas}. One of the main results of our work concerns the effect of such decoherence on the present chaotic bipartite system. In general, it has been shown that for quantum chaotic systems that possess a clear classical limit, decoherence is a possible candidate to explain how chaos, in a classical sense, emerges in quantum mechanics. In particular, decoherence can prevent rapid spreading of an initially localized state and the quantum-classical correspondence remains valid for long times~\cite{decochaos}. However, when the system does not support a classical limit, like here, the situation becomes more complex~\cite{bipartite}. 

In the present model the effects of quantum fluctuations in the two-level system are substantial and we cannot ascribe it a direct classical counterpart. As a consequence, we will demonstrate that the semi-classical predictions are not recovered by introducing measurement induced decoherence of the oscillator subsystem (electromagnetic field) in our model. We attempt to understand this lack of correspondence by showing that fundamentally non-classical characteristics, such as qubit-field entanglement, survive the decoherence. In the optical regime, the effects of a weak coupling to an environment or a measurement device is conveniently captured in terms of a master equation~\cite{gardiner,open}. In an unraveling approach to solving the master equation, the {\it Lindblad terms} constitute imaginary state decay together with stochastic {\it quantum jumps}. Averaging over all these quantum trajectories results in the evolved state of the system. Single stochastic trajectories, which can be interpreted as single experimental measurement realizations, are shown to support localization. As will be explained further below, it is important to distinguish this localization from dynamical localization \cite{haake,moore94,chaosloc,stöckmann} which is a quantum coherent effect that is destroyed by decoherence not generated by it \cite{klappauf98}. The localization phenomenon we discuss here only appears in the states of single stochastic trajectories and not in the full quantum probability distribution (i.e.\ the average over an ensemble of measurements). We argue that signatures of chaotic behaviour in circuit QED can be seen in the photon statistics of the transmission line output field.

This paper is organized as follows. The next section introduces the model system and the decoherence stemming from the position measurement. In section~\ref{sec3} we derive a set of semi-classical equations of motion where the field variable is treated at a mean-field level while a quantum treatment is maintained for the qubit. The results of the full quantum model are presented in section~\ref{sec4}. We especially focus on dynamics in phase space which can be visualized using Wigner and Husimi distributions, including localization appearing due to `{\it selective measurements}', and methods to detect signatures of quantum chaos. Finally, we conclude in section~\ref{sec5}. In the appendix we present pictures of typical examples of the residuals of the phase space probability distributions in order to give a better idea of how the individual $x$ and $p$ probability distributions look under the combined effects of unitary evolution and measurement.

\section{Physical model}\label{sec2}
Our physical model consists of a superconducting qubit coupled to a driven transmission line resonator as, for example, in the experiments described in references \cite{circuitQED} and \cite{circuitexp}. The undriven setup is described by the Rabi Hamiltonian~\cite{jonas1} ($\hbar=1$)
\begin{equation}\label{rabiham}
\hat{H}_R=\omega\left(\frac{\hat{p}^2}{2}+\frac{\hat{x}^2}{2}\right)+\frac{\Omega}{2}\hat{\sigma}_z+\sqrt{2}g\hat{x}\hat{\sigma}_x,
\end{equation}
where $\omega$ and $\Omega$ are the mode and qubit transition frequencies respectively, $g$ the effective qubit-light interaction strength, $\hat{x}$ and $\hat{p}$ the two field quadratures of the resonator mode, and $\hat{\sigma}_x=|2\rangle\langle1|+|1\rangle\langle2|$ and $\hat{\sigma}_z=|2\rangle\langle2|-|1\rangle\langle1|$ are the Pauli matrices acting on the two qubit states $|1\rangle$ and $|2\rangle$. We will work with dimensionless variables where $\hbar\omega$ sets the energy scale, i.e.\ from now on $\omega=1$. The above Hamiltonian can be microscopically derived from a minimal coupling model where the two-level, single mode, and dipole approximations are imposed, and the ``self-energy'' of the field is neglected, see Ref.~\cite{micro2}. In the rotating wave  approximation (RWA) one recovers the celebrated Jaynes-Cummings model~\cite{jc,jonas1}. In traditional microwave and optical cavity QED setups, $g\ll1$ such that this approximation is well justified~\cite{micro}, while recent advances in circuit QED make it possible to reach regimes where the RWA breaks down~\cite{usc}. When one goes beyond the RWA limit, especially for the ultrastrong coupling regime $g>\sqrt{\Omega}/2$, the photon population $\bar{n}_0$ of the system ground state increases rapidly ($\bar{n}_0\sim g^2$). For the traditional configuration, namely a two-level system interacting with a single resonator mode, this build-up of photon population $\bar{n}_0$ in the ground state is mainly due to the fact that the electromagnetic self-energy in the Hamiltonian (\ref{rabiham}) has been left out~\cite{nogo}. This problem can be circumvented by following the timely suggestion made in reference~\cite{solano1}: by adding an external drive for the qubit one obtains an effective model of the Rabi form, but where the qubit-field coupling is scaled with the pump amplitude while the self-energy is unaffected, and, therefore,  the two terms can be tuned independently. Hence, the self-energy can be neglected even in the ultrastrong coupling regime of the effective Rabi model. This fact is important for us here because we find that chaotic structures appear most definitively in the ultrastrong coupling regime.

Making use of the $Z_2$ parity symmetry of the Rabi Hamiltonian, the model was recently shown to be {\it quasi-solvable} over its entire parameter range~\cite{rabi_sol}. This integrability is somewhat surprising since the model does not possess a continuous symmetry. Moreover, integrability typically implies absence of chaos and the dynamics is therefore regular for all energies and parameters~\cite{jonasrabi}. However, as we will demonstrate, when the system is externally driven it may become chaotic for certain time-dependent drive amplitudes. The driven Rabi Hamiltonian is given by
\begin{equation}\label{rabiham2}
\begin{array}{lll}
\hat{H}_{dR} & = & \displaystyle{\hat{a}^\dagger\hat{a}+\frac{\Omega}{2}\hat{\sigma}_z+g\left(\hat{a}^\dagger+\hat{a}\right)\hat{\sigma}_x}\\ \\
& & +\eta(t)\left(\hat{a}^\dagger e^{-i\omega_d t}+\hat{a}e^{i\omega_dt}\right),
\end{array}
\end{equation}
where we have introduced the photon annihilation and creation operators  $\hat{a}=(\hat{x}+i\hat{p})/\sqrt{2}$ and $\hat{a}^\dagger=(\hat{x}-i\hat{p})/\sqrt{2}$, respectively. The drive amplitude will be taken to be 
\begin{equation}
\eta(t)=\eta_0\cos(\omega_ct),
\end{equation}
 i.e.\ the \emph{amplitude} is modulated in time at angular frequency $\omega_c$ between the maximum/minimum values of $\pm \eta_0$. The amplitude modulation frequency should not be confused with the central frequency $\omega_d$ of the drive field upon which it imposes side bands. Returning to the quadrature representation, we have
\begin{equation}\label{rabiham3}
\begin{array}{lll}
\hat{H}_{dR} & = & \displaystyle{\frac{1}{2}\left(\hat{p}+\sqrt{2}\eta(t)\sin(\omega_dt)\right)^2+\frac{\hat{x}^2}{2}+\frac{\Omega}{2}\hat{\sigma}_z}\\ \\
& & \displaystyle{\left(\sqrt{2}g\hat{\sigma}_x+\sqrt{2}\eta(t)\cos(\omega_dt)\right)\hat{x}-\frac{\eta^2(t)\sin^2(\omega_dt)}{2}},
\end{array}
\end{equation}
which will be our starting Hamiltonian. 

To gain further insight, it is convenient to perform an adiabatic diagonalization where the Hamiltonian is expressed in the eigenvalue basis of the two-level matrix $V(x)=\Omega\hat{\sigma}_z/2+gx\hat{\sigma}_x$ with corresponding adiabatic potentials
\begin{equation}
V_{ad}^\pm(x,t)=\frac{x^2}{2}+\sqrt{2}\eta(t)\cos(\omega_dt)x\pm\sqrt{\frac{\Omega^2}{4}+2g^2x^2}.
\end{equation}
The Born-Oppenheimer approximation consists of neglecting any coupling between the resulting adiabatic eigenstates~\cite{BO}. If we let $\eta_0=0$ we note that the lower potential $V_{ad}^\pm(x)$ has a double-well structure whenever $g>\sqrt{\Omega}/2$. The driving induces a back and forth rocking of the double-well. This type of driven double-well system mimics the Duffing oscillator~\cite{duffing}. However, for non-zero $\omega_d$ the driving also induces a ``shaking'' of the momentum. From Eq.~(\ref{rabiham3}) it follows that this shaking effect can be accorded a time-dependent gauge potential $\hat{A}(t)=\sqrt{2}\eta(t)\sin(\omega_dt)$ giving rise to a synthetic electric field $\hat{E}=-\partial_t\hat{A}(t)$. While interesting, in this paper we shall not discuss this gauge aspect further. As will be described in the next section, the phenomena we are interested in stem from the rocking of the potential and this synthetic electric field will not change our arguments.  

Finally we want to include the impact of quadrature measurements~\cite{duffingmeas}. In the Schr\"odinger picture and within the {\it Born-Markov approximation} a standard approach for describing decoherence is given by considering a master equation where the irreversible processes are attributed to a set of {\it Lindblad operators}~\cite{open}. Instead of considering field and qubit decay, we consider decoherence arising from the measurement of the $\hat{x}$ quadrature of the field. Thus, we work with a Markovian master equation of the form
\begin{equation}\label{mastereq}
\frac{d}{dt}\hat{\rho}=i[\hat{\rho},\hat{H}_{dR}]+\mathcal{L}[\hat{x}]\hat{\rho},
\end{equation}
where $\hat{\rho}$ is the density operator for the full system, and the Lindblad operator
\begin{equation}\label{lindbladop}
\mathcal{L}[\hat{x}]\hat{\rho}=\kappa\left(2\hat{x}\hat{\rho}\hat{x}^\dagger-\hat{x}^2\hat{\rho}-\hat{\rho}\hat{x}^2\right)=-\kappa\big[\hat{x},\left[\hat{x},\hat{\rho}\right]\big]
\end{equation}
accounts for the irreversible loss of coherence due to the measurement of the quadrature $\hat{x}$~\cite{duffingmeas}. The information loss rate is given by the parameter $\kappa$. Intuitively, a large $\kappa$ is expected to produce a localization effect upon the quantum state. However, if the measurement back-action becomes too large it will inevitably affect the variable conjugate to the measured one. Thus, in order to achieve a localization effect, $\kappa$ should be in between the two extremes. In the present bipartite model this picture will, however, be very different. Note that the form of $\mathcal{L}[\hat{x}]$ implies only loss of coherence and no dissipation. This will allow us to identify any localization effect in phase space as being due to decoherence rather than a reduction of the available phase space due to energy loss. We point out that our results are not unique to decoherence of the form of (\ref{lindbladop}) but other decoherence channels like $\mathcal{L}[\hat{a}^\dagger\hat{a}]$ give similar behaviour (as verified numerically). For pure photon decay $\mathcal{L}[\hat{a}]$ which also induces dissipation, the quantum-classical crossover in a Jaynes-Cummings model with a Kerr medium was studied in Ref.~\cite{ev1} in terms of collapse-revivals.  

A further remark before proceeding: whenever the system at hand is the subject of decoherence, information about it is irreversibly lost leading to an effective non-linear theory that can become chaotic in a classical sense, i.e. for short times nearby initial states move apart exponentially with a rate set by the maximum {\it Lyapunov exponent}~\cite{strogatz}. However, in an analysis like ours, any irreversible process in quantum mechanics derives from the effective nature of the theory. Normally, it is assumed that the measurement device is macroscopic so that information about the system is distributed over a large number of degrees of freedom and a reversed flow of information back into the system is very unlikely. This justifies the Markovian assumption of infinitely fast decay of correlations within the measurement device~\cite{milburn,open}. It is nevertheless an approximation and by waiting long enough information will flow back into the system (quantum recurrences). This waiting time will typically be very long in comparison to any experimentally relevant time scale so that the physics will appear to be non-linear in the experiment. 

\section{Semi-classical analysis}\label{sec3}
At large pumping, $\eta_0>1$, we replace the mode variables by their corresponding $c$-numbers as obtained in a coherent state ansatz. Then, the resulting effective Hamiltonian acts on the qubit sub-space alone, while its parameters depend on the boson field and must be determined self-consistently. In such a mean-field approach we neglect any quantum correlations between the qubit and the field, i.e.\ the qubit remains in a pure state and its presence does not imply decoherence of the large amplitude resonator field. This allows us to derive a closed set of equations of motion, and thereby characterize the underlying semi-classical dynamics, i.e.\ whether it is regular, chaotic, or mixed.

Since the qubit is decoupled from the field, we can describe it via two variables. After denoting the qubit state as $|\chi\rangle=\left[\beta_1e^{-i\phi_1}\,,\,\beta_2e^{-i\phi_2}\right]^T$, where $\beta_j$ and $\phi_j$ ($j=1,\,2$) are real, we introduce the (conjugate variables) {\it inversion} $Z=\beta_1^2-\beta_2^2$ and relative phase $\Delta_\phi=\phi_1-\phi_2$. The semi-classical equations of motion become
\begin{equation}\label{eom}
\begin{array}{lll}
\displaystyle{\dot{x}= p+\sqrt{2}\eta(t)\sin(\omega_dt),}\\ \\
\displaystyle{\dot{p}=- x-g\sqrt{2}\sqrt{1\!-Z^2}\cos(\Delta_\phi)-\sqrt{2}\eta(t)\cos(\omega_dt),}\\ \\
\dot{Z}=g\sqrt{2}x\sqrt{1-Z^2}\sin(\Delta_\phi),\\ \\
\displaystyle{\dot{\Delta}_\phi=\frac{\Omega}{2}-g\sqrt{2}x\cos(\Delta_\phi)\frac{Z}{\sqrt{1-Z^2}}},
\end{array}
\end{equation}
where the dot denotes a time-derivative. We have left out the decoherence induced Langevin forces~\cite{gardiner} in this mean-field analysis. As a result, the semi-classical dynamics is independent of $\kappa$ which reflects the fact that the quadrature measurement only induces decoherence and no dissipation. From Eqns~(\ref{eom}), we find the semi-classical Hamiltonian
\begin{equation}
\begin{array}{lll}
H_{scl}(x,p,Z,\Delta_\phi) & = & \displaystyle{\frac{p^2}{2}+\frac{x^2}{2}+p\sqrt{2}\eta(t)\sin(\omega_dt)}\\ \\
& & \displaystyle{+x\sqrt{2}\eta(t)\cos(\omega_dt)}\\ \\ & & \displaystyle{+gx\sqrt{2}\sqrt{1-Z^2}\cos(\Delta_\phi)+\frac{\Omega}{2}Z}.
\end{array}
\end{equation}
We stress that the mean-field approximation is only applied to the boson degree-of-freedom, in Eq.~(\ref{eom}) the spin is treated quantum mechanically. 

From Eq.~(\ref{eom}) we directly identify a trivial stable fixed point given by $x_{ss}=p_{ss}=0$ whenever $\eta_0=0$ (i.e. undriven system). At $g^2=\Omega/4$, there is a Pitchfork bifurcation where the fixed point turns unstable and two new stable fixed points arise ($g^2>\Omega/4$), $x_{ss}=\pm g\sqrt{2-\frac{\Omega^2}{8g^4}}$. These correspond to the minima of the double-well potential $V_{ad}^-(x)$. In an adiabatic picture, the driving induces an effective time-dependent ``detuning'' between the two double-well minima. For moderate driving, this dynamics can be treated as repeated Landau-Zener transitions~\cite{lz}. For large pumping $\eta_0$ (typically $|\eta_0|>|g|$), however, the double-well structure may be lost at the extreme of the driving, i.e. when $|\eta(t)|$ attains its maximum. In this case, the lower adiabatic potential oscillates between having one or two local minima.  

\begin{figure}[h]
\centerline{\includegraphics[width=8cm]{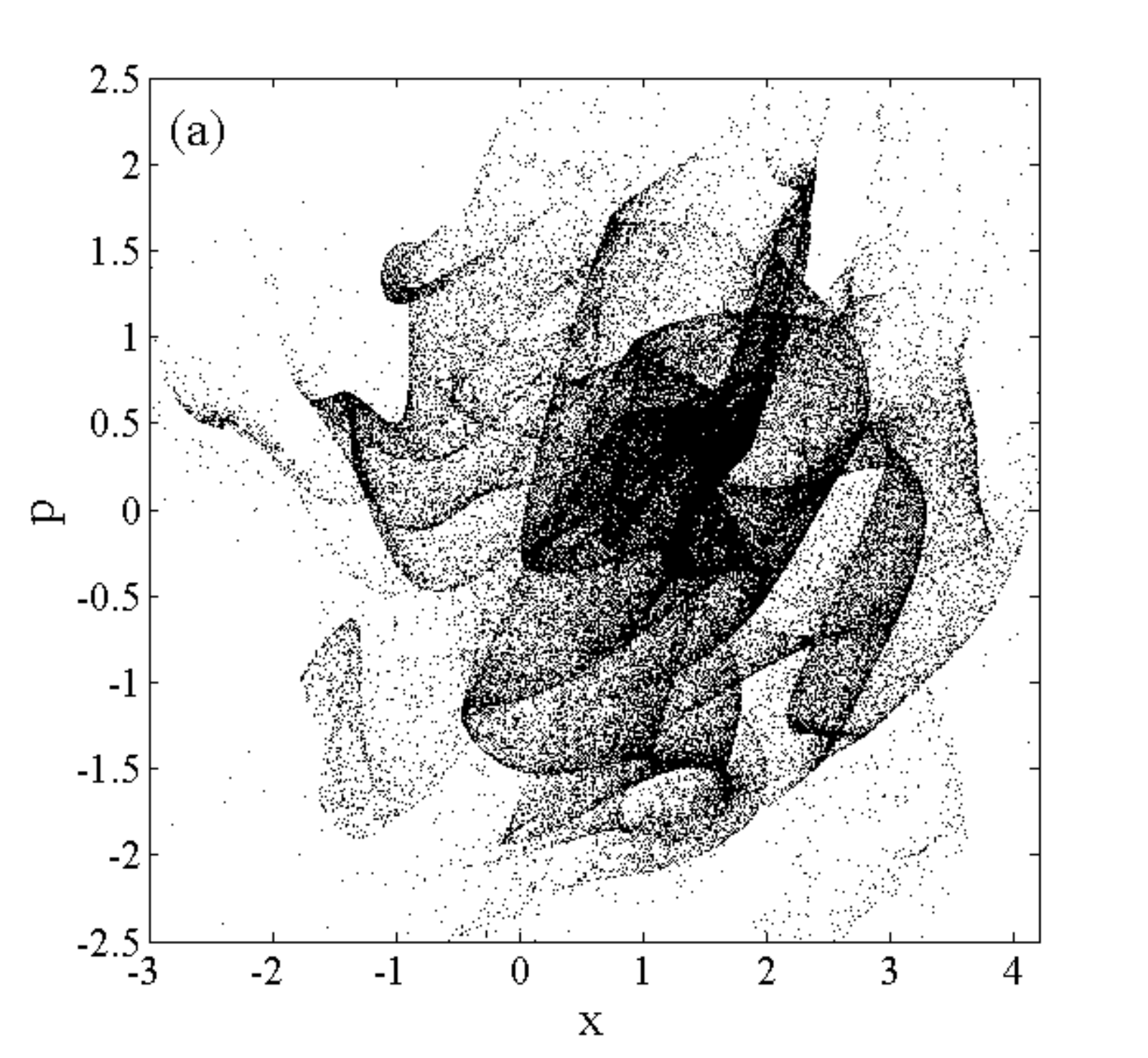}}
\centerline{\includegraphics[width=8cm]{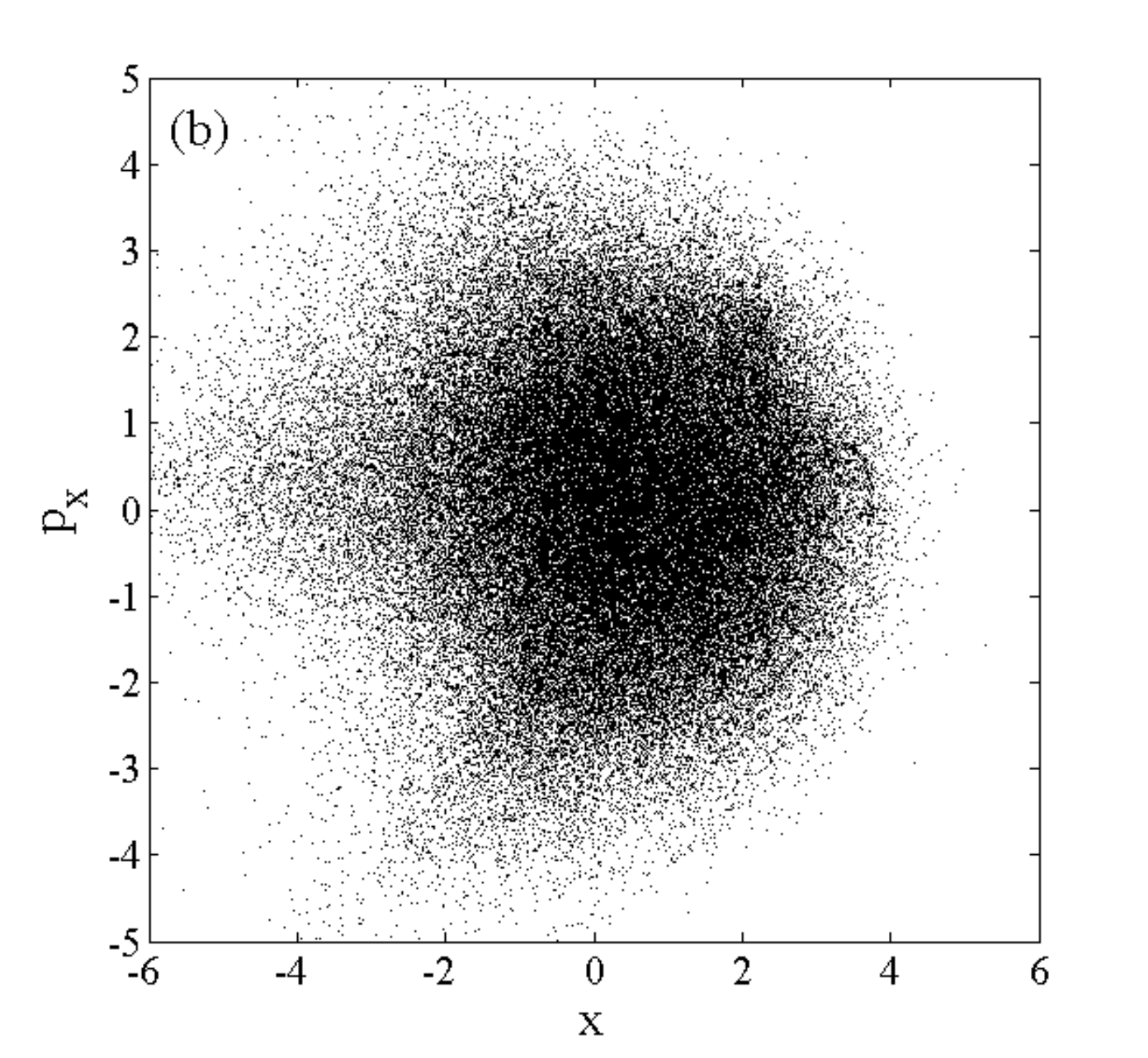}}
\caption{Stroboscopic maps at $t=T$ (a) and $t=4T$ (b). In (a), after one period $T$, the field phase space distribution still displays some structures, while in (b) they have more or less fully disappeared. The dimensionless parameters are $\Omega=\omega_d=1$, $\omega_c=g=1.5$, and $\eta_0=3$. The number of simulated trajectories $N_{cl}=100\,000$. } \label{fig1}
\end{figure}

In this paper we use the semi-classical equations of motion to perform an analysis of the dynamics within the {\it truncated Wigner approach} (TWA)~\cite{twa}. That is, we simulate $N_{cl}\gg1$ trajectories of the above Eqns~(\ref{eom}) with initial conditions for $x$ and $p$ randomly picked according to a Gaussian probability distribution corresponding to a coherent state $|\alpha\rangle$ with amplitude $\alpha=x_{ss}/\sqrt{2}$. The qubit initial state is taken $\Delta_\phi(t=0)=\pi$ and $Z(t=0)=0$. This implies that the initial state predominantly populates the right minimum of the lower adiabatic potential $V_{ad}^-(x)$. The set of differential equations~(\ref{eom}) are solved using the 4th order Runge-Kutta algorithm as modified by {\it Gear's method}, which is especially suited to stiff equations.

Throughout this work we will consider the resonant situation $\Omega=\omega_d=\omega=1$, the pump amplitude $\eta_0=3$, amplitude modulation frequency $\omega_c=1.5$, and effective atom field coupling $g=1.5$. This choice means that we are well into the ultrastrong coupling regime $g>\sqrt{\Omega}/2$, and, as we shall see below, $\eta_0=3$ is large enough to guarantee classical chaos. It should be pointed out that the results are not sensitive to the particular parameter values as long as the dynamics is chaotic. In fact, by fine tuning the drive parameters, chaotic evolution is also attainable in the ``normal phase'' $g<\sqrt{\Omega}/2$ where the lower adiabatic potential has a global minimum at $x=0$. 

With $\omega_c=3\omega_d/2$, the Hamiltonian is periodic with period $T=4\pi$. We consider {\it stroboscopic maps} of the phase space distribution of the resonator mode at times $t=T$ and $4T$. As can be seen in Fig.~\ref{fig1}, after one period the distribution still shows some visible structures, while after four periods almost all such structures are washed out. For this choice of initial state and system parameters, the dynamics is completely chaotic lacking any islands of regular motion~\cite{strogatz}. 

For a time-independent system, chaotic dynamics typically smears out the distribution over the accessible phase space energy shells. We note that the corresponding effect in our time-dependent system occurs at rather short time scales, e.g.\ the classical period of the harmonic oscillator $T_{cl}=2\pi$ in scaled dimensionless units. It is clear that the stroboscopic maps of Fig.~\ref{fig1} show different characteristics compared to those of the quantum Duffing oscillator~\cite{duffing_classlim}. This results from the additional degree-of-freedom of the spin. The effective phase space is here four dimensional instead of two dimensional as it is for the Duffing oscillator and therefore the typical stretching and squeezing structure of the distribution~\cite{duffing_classlim} is not visible in the stroboscopic maps shown in Fig.~\ref{fig1} since they are projections onto the two dimensional $xp$-plane. 

\section{Quantum analysis}\label{sec4}
The master equation (\ref{mastereq}) gives the evolution of the quantum state $\hat{\rho}$ under the influence of a certain form of decoherence which can be interpreted as a continuous position measurement. Extracting $\langle\hat{x}\rangle$ implies measuring the field quadrature for an ensemble of equally prepared initial states~\cite{micro}. The statistical average of such homodyne measurement sequences gives $\langle\hat{x}\rangle$. Similarly, when solving (\ref{mastereq}) numerically it is practical to transform the problem into one of solving a sequence of independent stochastic Schr\"odinger equations, each one of them resulting in a solution $|\psi_i(t)\rangle$ (note that this is a pure state). In this unraveling method any expectation value becomes $\langle\hat{A}\rangle=\mathrm{Tr}\left[\hat{A}\hat{\rho}(t)\right]=\lim_{N\rightarrow\infty}\frac{1}{N}\sum_i^N\langle\psi_i(t)|\hat{A}|\psi_i(t)\rangle$. Each $\langle\hat{A}\rangle_i=\langle\psi_i(t)|\hat{A}|\psi_i(t)\rangle$ can be viewed as one of the measurement sequences and every photon detection (a `click' in the detector) is represented by a single `quantum jump' in the stochastic Schr\"odinger equation. It is important to appreciate the distinction between the outcome from a single measurement sequence and one deriving from an ensemble of measurements. We note that both are of experimental relevance.  

For the numerics we employed the {\it split-operator method}~\cite{split} combined with the {\it quantum Monte Carlo method}~\cite{qmc}. The split-operator method relies on factorizing, for short times $\delta t$, the evolution operator into a real-space and a momentum part. The corresponding two propagators are applied in real and momentum space respectively. The method becomes exact in the limit $\delta t\rightarrow0$. In general, the inherent time scales set the constraints on the size of $\delta t$. The final time is $t=K \delta t$, where $K$ the number of steps used in the propagation. We varied $\delta t$ in order to check the convergence of our extracted results. The split-operator method allows us to simulate the coherent evolution, while for decoherence we apply the quantum Monte Carlo method, in which $N$ different trajectories $|\psi_i(t)\rangle$ are simulated and where for each simulation $i$ we periodically interrupt the coherent propagation and randomly choose either the absence or occurrence of a quantum jump characterizing a photon detection~\cite{micro}. This interruption is done $K$ times, i.e. it agrees with the time-steps used for the coherent propagation, and we have verified that the jump probability $P_{jump}\ll1$ in each step. Note that a quantum jump in the present homodyne scheme is represented by multiplying the state $|\psi(t)\rangle$ by $\hat{x}$ (and not by $\hat{a}$ as in a normal detection scheme where the photon intensity leaking out of the cavity is measured). As an initial state we will consider a coherent state predominantly occupying one of the two minima of $V_{ad}^-(x,t=0)$. Such a state can be approximated as $|\psi_\pm(t=0)\rangle=|\pm\alpha\rangle|\Theta_\pm\rangle$, where $\alpha=x_{ss}/\sqrt{2}$ is a real coherent state amplitude and $|\Theta_\pm\rangle=[\pm1\,,\,1]^T/\sqrt{2}$ is the initial qubit state. The $\pm$-sign represents the left or right potential minimum, respectively. This state approximates the ones used for the initialization of the TWA considered in the previous section. Notice that the initial state represents a disentangled field-qubit state.

\begin{figure}[h]
\centerline{\includegraphics[width=8cm]{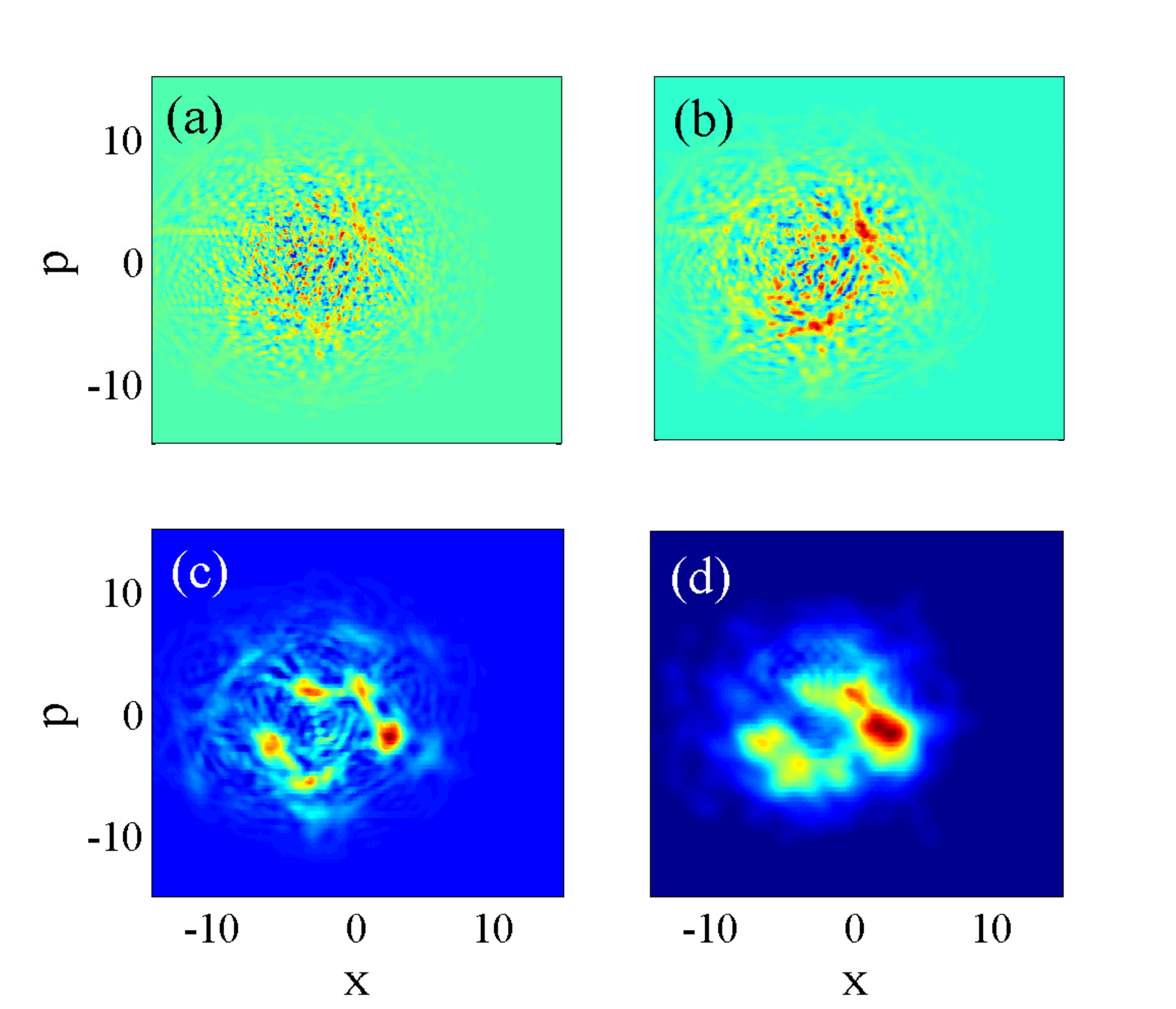}}
\caption{(Color online) The Wigner phase space distribution~(\ref{wigner}) for the resonator field and for four different decay rates $\kappa=0,\,0.005,\,0.01,\,0.05$ (a)-(d) respectively. The effect of decoherence is seen to have several consequences; $i$ overall smoothening - washing out sub-Planck structures, $ii$ slight suppressing of phase space spreading, and $iii$ abate negativity. The rest of the dimensionless parameters are as in Fig.~\ref{fig1}, with a propagation time $t=200$ for all four plots. } \label{fig3}
\end{figure}

\begin{figure}[h]
\centerline{\includegraphics[width=8cm]{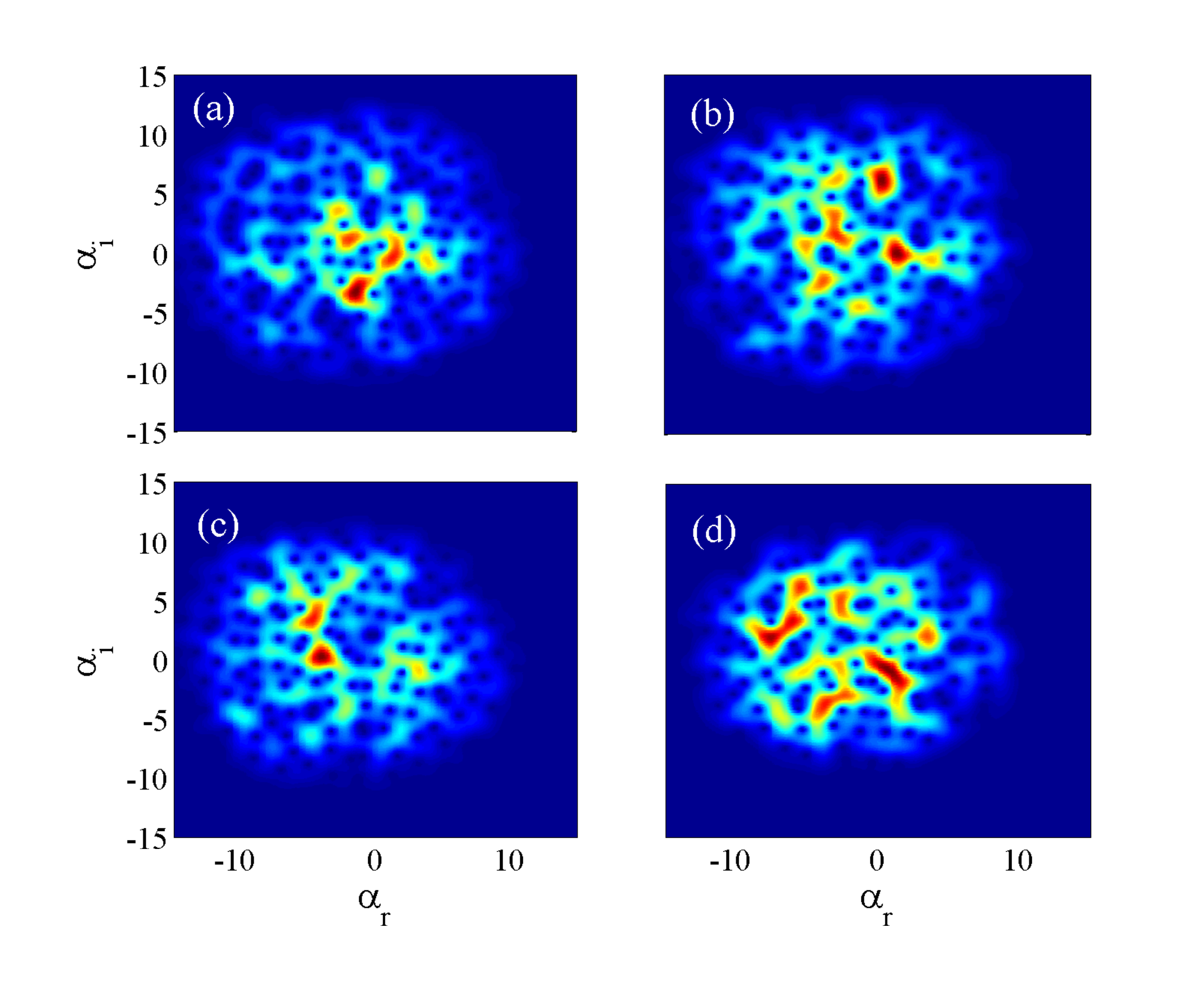}}
\caption{(Color online) Identical to Fig.~\ref{fig3} but displaying the Husimi distribution~(\ref{husimi}) instead. In contrast to the Wigner distribution, the Husimi distribution is positive definite and lacks sub-Planck scale structures due to the Gaussian averaging as discussed in the main text. } \label{fig4}
\end{figure}

A Markovian master equation of the Lindblad form may imply (depending on the actual form of the Lindblad terms) dissipation and decoherence. In a driven system, dissipation may balance the pumping such that a steady state containing a finite average number of photons is approached~\cite{gardiner}. The main result of decoherence, on the other hand, is to suppress the off-diagonal terms of the density operator $\hat{\rho}$~\cite{phoenix}. This diagonalization can hinder the evolution, similar to the {\it quantum Zeno effect}. For chaotic systems, the role of reservoir induced decoherence has been discussed in numerous works, see for example~\cite{duffing_classlim,disschaos,brun,decochaos}. Decoherence stemming from quantum measurements has been explored as well, see Ref.~\cite{duffingmeas,decomeas}. In the case of the quantum Duffing oscillator, it has been verified that the expectation values $\langle\hat{A}\rangle_{t}$ obtained from the corresponding classical model agreed much better with those of the quantum model when decoherence is included than those given by the closed quantum system~\cite{duffing_classlim}. Moreover, the expectation values can become sensitive to small fluctuations in the initial states when decoherence is included, a typical characteristic for chaotic dynamics. At the same time, it has long been known that decoherence is the main source for loss of `quantumness', e.g.\ quantum entanglement or negativity of the Wigner function~\cite{milburn,open}. It might therefore be expected that decoherence helps in obtaining a proper quantum-classical correspondence for chaotic systems. While the issue of decoherence in chaotic systems has been rather well studied for cases in which the system has an obvious classical limit, for system with no clear classical limit little is known~\cite{bipartite}. The situation is even more complex in bi- or multi-partite systems where entanglement may persist between the different subsystems even in the presence of decoherence~\cite{knight}. 

\begin{figure}[h]
\centerline{\includegraphics[width=8cm]{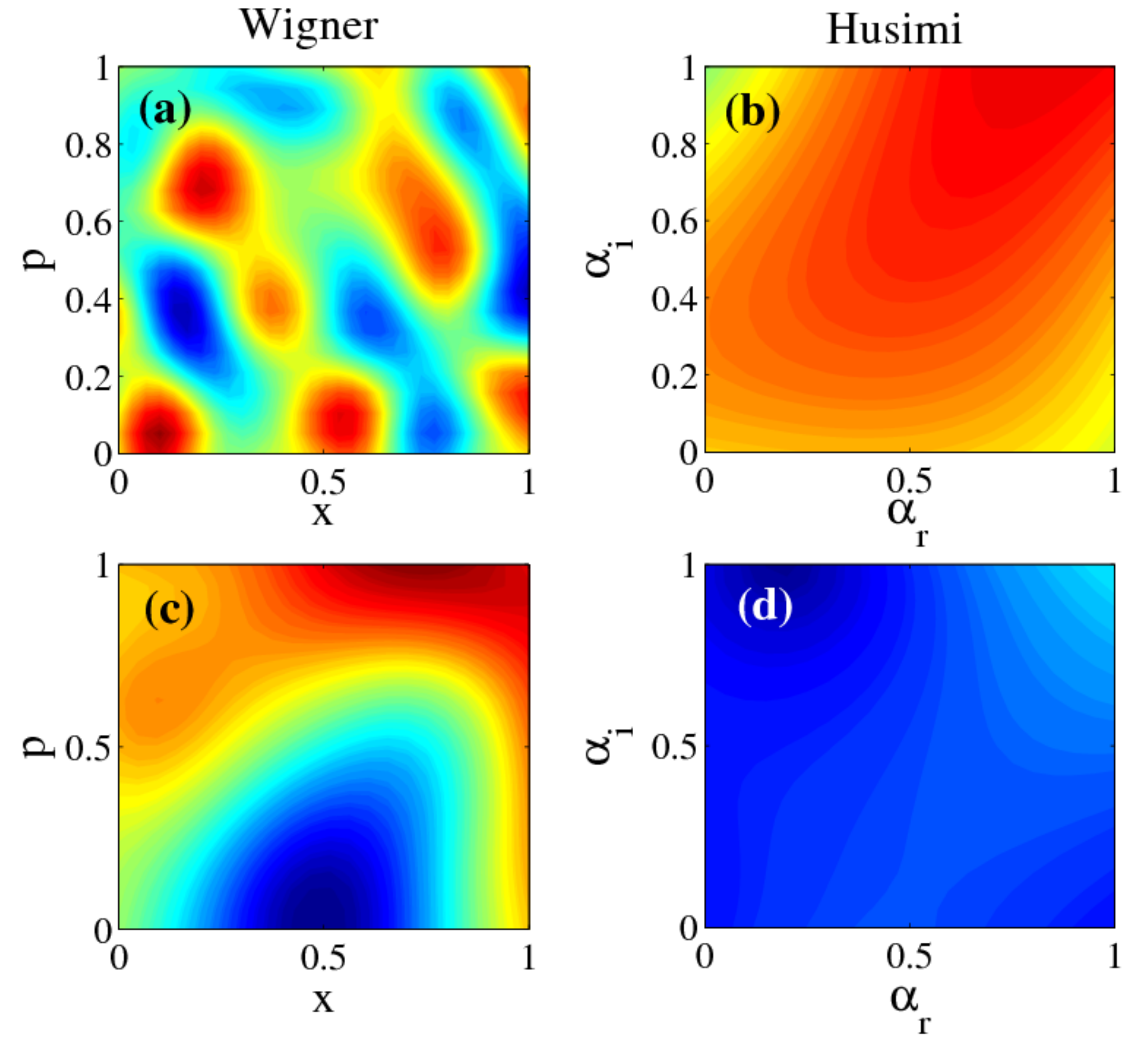}}
\caption{(Color online) One Planck cell of the Wigner distribution (left) and the Husimi distribution (right). For the upper two plots, (a) and (b), the system is closed ($\kappa=0$) and the Wigner distribution contains structures smaller than the Planck cell. In (c) and (d), on the other hand, decoherence ($\kappa=0.01$) smoothens the Wigner distribution and thereby prevents fine structures from forming. The other parameters are as in Figs.~\ref{fig3} and \ref{fig4}.} \label{fig4b}
\end{figure} 

\subsubsection{Phase space analysis}
Quantum phase space distributions are a practical tool for achieving a deeper intuition for quantum dynamical systems~\cite{scully,gardiner}. The idea is to visualize the quantum evolution in a phase space picture. The Wigner distribution for a quantum state $\hat{\rho}(t)$ (in our case the resonator boson mode) is defined as~\cite{scully}
\begin{equation}\label{wigner}
W(x,p,t)=\frac{1}{\pi}\int dy\,\langle x-y/2|\hat{\rho}(t)|x+y/2\rangle\, e^{ipy}.
\end{equation}
Because it does not have to be positive everywhere in phase space, $W(x,p,t)$ is not a proper probability distribution. Typically, negativity of $W(x,p,t)$ is identified with non-classical phenomena such as quantum coherent superpositions. The Wigner distribution is normed to unity, $\iint dxdp\, W(x,p,t)=N_-(t)+N_+(t)=1$, where $N_\pm(t)$ measures the negative ($-$) and positive fraction ($+$) of the distribution. Note that normalization does not imply $|N_\pm(t)|<1$. The marginal distributions of $W(x,p,t)$ give the real and momentum space probability distributions, i.e. they are positive definite. Another characteristic of $W(x,p,t)$ is that it allows for sub-Planck structures~\cite{zurek2} which is another indication that the Wigner function cannot be regarded a formal probability distribution. As a counterpart to classical phase space distributions, the Husimi $Q$-function seems more of a natural choice. The $Q$-function is positive definite and in contrast to the Glauber $P$-functions, it does not become singular. It is defined as~\cite{scully,gardiner} 
\begin{equation}\label{husimi}
Q(\tilde{x},\tilde{p},t)=\frac{1}{\pi}\langle\alpha|\hat{\rho}(t)|\alpha\rangle,
\end{equation}
where $|\alpha\rangle$ is a coherent state with amplitude $\alpha=\alpha_r+i\alpha_i\equiv\sqrt{2}(\tilde{x}+i\tilde{p})$ ($\alpha_r$ and $\alpha_i$ are hence both real). It follows that the Husimi $Q$-function is a Gaussian averaged Wigner function~\cite{rajagopal},
\begin{equation}
Q(\tilde{x},\tilde{p},t)=\frac{1}{\pi}\iint dxdp\,e^{-(x-\tilde{x})^2-(p-\tilde{p})^2}W(x,p,t).
\end{equation}
This smoothening not only extinguishes the negativity, but also prevents sub-Planck structures from forming~\cite{taka}. Interestingly, despite this averaging, because of the overcompleteness of coherent states the $Q$-function still contains all the information there is to know about the state $\hat{\rho}(t)$. While the $Q$-function has the appealing property of being positive definite, its marginal distributions do not reproduce the correct real and momentum space probability distributions. Note that the phase space distributions are defined for general states $\hat{\rho}(t)$, with no restriction on purity. This applies especially to us where we are interested in the distributions for the field. The density operator for the field $\hat{\rho}_f(t)$ is typically in a mixed state even without coupling to the environment since it is obtained from the full density operator by a trace over the qubit's degrees-of-freedom, $\hat{\rho}_f(t)=\mathrm{Tr}_\mathrm{qu}\left[\hat{\rho}(t)\right]$. 

For a chaotic system like the present one, we expect the phase space distributions to show an ergodic collapse in which they spread out over the accessible phase space. Even at resonant driving, $\omega_d=\Omega=1$, the model appears to have a relatively bounded energy shell due to the oscillating pump amplitude $\eta(t)$ and the coupling to the qubit. After the collapse, a seemingly irregular phase space structure builds up. In a closed quantum system, the above behaviour is typical for quantum thermalization~\cite{jonas4}. How the phase space structures develop in a chaotic open system is, however, not well explored. In our model we cannot talk about {\it quantum thermalization} since the system is both driven and exposed to decoherence. 

\begin{figure}[h]
\centerline{\includegraphics[width=8cm]{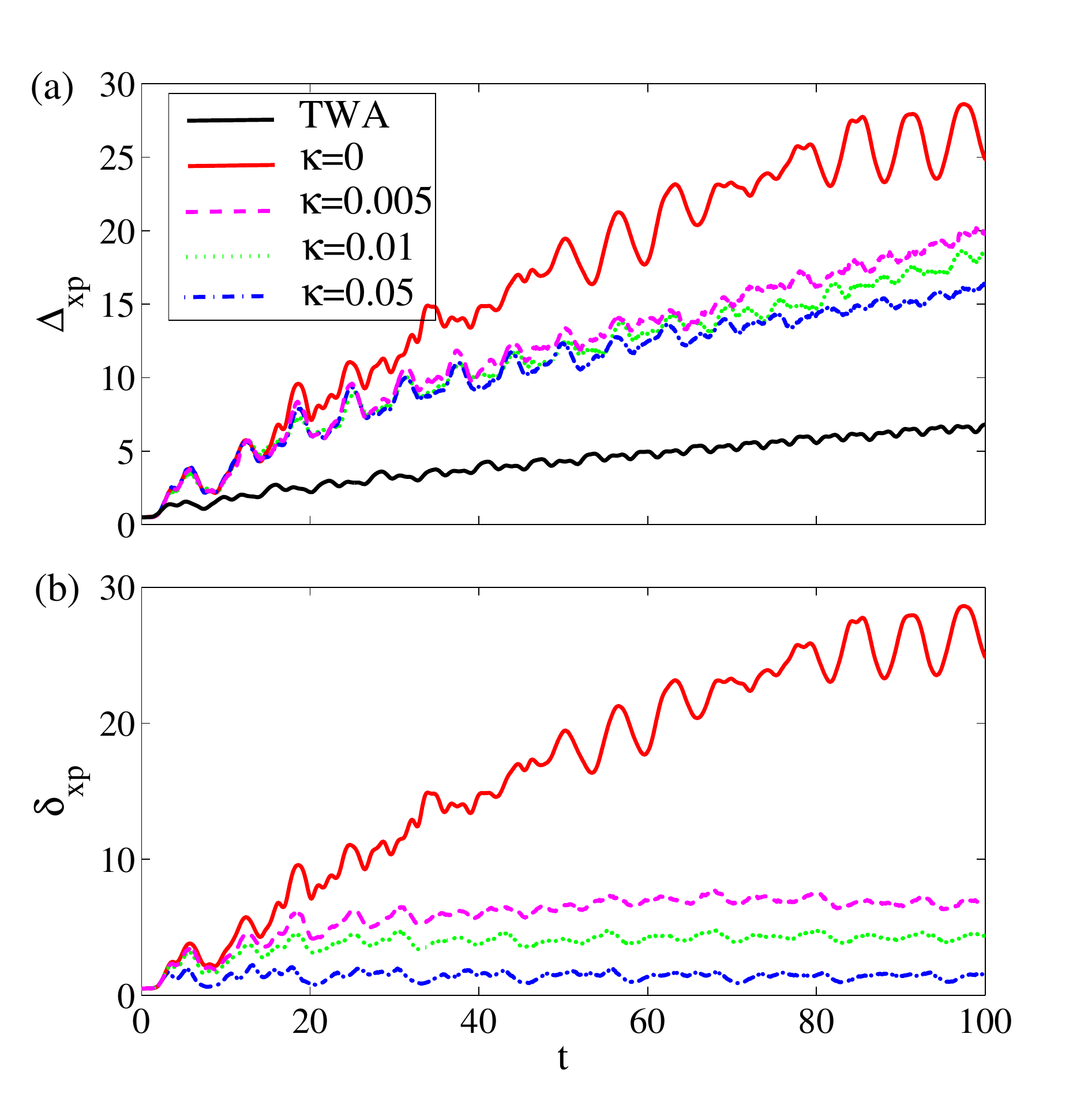}}
\caption{(Color online) Evolution of the effective phase space widths $\Delta_{xp}$ (a) and $\delta_{xp}$ (b) for various decay rates. The black line in the upper plot is obtained from the semi-classical TWA result. The remaining lines (solid red, dot-dashed blue, dotted green, and dashed purple) display the full quantum results. The width $\Delta_{xp}$ of the full phase space distribution shows a weak $\kappa$-dependence. In contrast, single trajectory widths $\delta_{xp}$ have a strong $\kappa$-dependence which demonstrates a localization effect. The rest of the dimensionless parameters are the same as in Fig.~\ref{fig1}.} \label{fig5}
\end{figure}

We have numerically calculated $W(x,p,t)$ and $Q(\tilde{x},\tilde{p},t)$ for different decay rates $\kappa$. The results are shown in Figs.~\ref{fig3} and \ref{fig4} respectively, while Fig.~\ref{fig4b} gives a zoom-in of a single Planck cell. Comparing the figures, it is apparent that the sub-Plank structures in the Wigner distribution ($\hbar=1$) do not occur in the $Q$-function. Furthermore, as expected, decoherence enters as a smoothening of the Wigner distribution, and for large $\kappa$ the sub-Planck structures are completely lost in the Wigner distribution. 

Instead of deriving the distributions directly from the state $\hat{\rho}_f(t)$, they can alternatively be obtained from the corresponding Fokker-Planck equations~\cite{gardiner}. This approach gives us additional insight into the phase space dynamics. The full bipartite $Q$-function~\cite{gardiner}, $Q(z,\alpha)=\frac{2}{\pi^2(1+|z|^2)}\langle z,\alpha|\hat{\rho}(t)|z,\alpha\rangle$, where $|z\rangle$ is a spin coherent state with complex amplitude $z$~\cite{spincoher}, obeys the equation
\begin{equation}
\partial_tQ(z,\alpha,t)=\left(\mathcal{L}+\mathcal{L}_\mathrm{qdiff}+\mathcal{L}_\mathrm{mdiff}\right)Q(z,\alpha,t).
\end{equation}
The first differential operator on the r.h.s represents a classical drift
\begin{equation}
\begin{array}{lll}
\mathcal{L} & = & \displaystyle{i\eta(t)e^{-i\omega_dt}\partial_\alpha+i\partial_\alpha\left[\alpha+3g\frac{z+z^*}{1+|z|^2}\right]}\\ \\ & & +i\partial_z\left[-\Omega z+g(1-z^2)(\alpha+\alpha^*)\right]+c.c.,
\end{array}
\end{equation}
the second quantum diffusion
\begin{equation}
\mathcal{L}_\mathrm{qdiff}=ig\partial_\alpha\partial_z(1-z^2)+c.c.,
\end{equation}
and the final term describes the measurement induced diffusion
\begin{equation}
\mathcal{L}_\mathrm{mdiff}=-\frac{\kappa}{2}\left(\partial_\alpha-\partial_{\alpha^*}\right)^2.
\end{equation}
From $Q(z,\alpha,t)$ we find the field $Q$-function by integrating out the spin coherent state variable $z$. The form of the second term $\mathcal{L}_\mathrm{qdiff}$ prevents sub-Planck structures from forming in agreement with the Heisenberg uncertainty relation~\cite{altland}, i.e.\ the origin of this term is the {\it quantum pressure} which becomes very large whenever the distribution is tightly squeezed. This explains the quantum nature of this effect (and this term indeed vanishes in the classical limit $\hbar\rightarrow0$). Note further that $\mathcal{L}_\mathrm{qdiff}$ mixes spin and field variables. The last term $\mathcal{L}_\mathrm{mdiff}$ stems from the noise resulting from the position measurement and only contains field variables. In this model, where $\hat{x}$ is being measured, it has a diffusive effect on its conjugate variable $\hat{p}$. Its form is the same in the Fokker-Planck equation obeyed by the Wigner distribution which explains the smoothening effect seen in Fig.~\ref{fig3} as $\kappa$ is increased. To summarize: $\mathcal{L}$ reproduces the classical dynamics, $\mathcal{L}_\mathrm{qdiff}$ gives quantum corrections, e.g. prevents sub-Planck structures from forming and generates qubit-field entanglement, and finally $\mathcal{L}_\mathrm{mdiff}$ describes the back-action of the measurement.

As already mentioned, chaotic evolution typically implies a rapid spreading of the phase space distribution. The `size' of the distribution can be estimated from the width
\begin{equation}\label{width1}
\Delta_{xp}=\Delta x\Delta p=\sqrt{\left(\langle\hat{x}^2\rangle-\langle\hat{x}\rangle^2\right)\left(\langle\hat{p}^2\rangle-\langle\hat{p}\rangle^2\right)},
\end{equation}
i.e.\ the product of the quadrature uncertainties. Figs.~\ref{fig3} and \ref{fig4} give the distributions as evolved according to the master equation (\ref{mastereq}), but as discussed above we may also consider the distributions $Q_i(\tilde{x},\tilde{p},t)$ and $W_i(x,p,t)$ for single trajectories $i$. Each one of these, corresponding to a single stochastic simulation of the unraveled master equation, gives a width $\delta_{xp}^{(i)}=\sqrt{\left(\langle\hat{x}^2\rangle_i-\langle\hat{x}\rangle_i^2\right)\left(\langle\hat{p}^2\rangle_i-\langle\hat{p}\rangle_i^2\right)}$ which is used to define the average width  
\begin{equation}\label{width2}
\delta_{xp}=\frac{1}{N}\sum_{i=1}^N\delta_{xp}^{(i)}.
\end{equation}     
$\Delta_{xp}$ and $\delta_{xp}$ represent, respectively, the {\it non-selective width} in which measurement results are not recorded and {\it selective width} in which the measurements are projective and information about the results of a measurement are 
fed back into the system. For the same decay rates $\kappa$ as in Figs.~\ref{fig3} and \ref{fig4}, we display the time-evolution of $\Delta_{xp}$ and $\delta_{xp}$ in Fig.~\ref{fig5} (a) and (b).  For a comparison to the semi-classical results we have also calculated $\Delta_{xp}$ within the TWA framework using (\ref{eom}).  In addition, in the appendix we have provided snap shots of the individual $x$- and $p$-quadrature probability distributions in order to show how they look. Fig.\ \ref{fig5} reveals numerous interesting phenomena:

\begin{enumerate}
\item Even for large decoherence rates $\kappa$, the position measurement does not localize the state $\hat{\rho}_f(t)$ such that it can be said to approximate the semi-classical result. This is in stark contrast with results found for the Duffing oscillator~\cite{duffingmeas}. In Ref.~\cite{bipartite} a similar breakdown of the quantum-classical correspondence was conjectured to be due to the quantumness of one of the  subsystems (two-level system). Below we will motivate this by calculating the qubit-field entanglement, and it will be shown that despite strong decoherence substantial non-classical correlations between the qubit and the field survives.  

\item A much more distinct localization effect is found in the selective width $\delta_{xp}$. For large times, the quantum distributions can even remain more localized than the semi-classical one given by the TWA. We have verified that the localization occurs in both conjugate variables $x$ and $p$, and not only in the measured position $x$ (see also the Appendix).


The localization here occurs due to quantum jumps. Every time a jump occurs, the wave function $\psi(x,t)$ is multiplied by $x$. Such a `measurement' interrupts the evolution and these stochastic kicks of the wave function cause a destructive interference leading to the localization effect. This is reminiscent of dynamical localization, which is the analogue in time of 1D Anderson localization in space, and occurs because of random (but coherent) interference that prevents the wave function from spreading (the latter has in fact been seen in a periodically ``kicked" optical cavity filled with a non-linear medium \cite{chaosqed}). Here the kicks result from measurements. However, it is interesting to note that the wave function of a single simulated trajectory $\psi_i(x,t)$ is always a pure state and so the mechanism for dynamical localization is perhaps quite close even if the origins are different. Meanwhile, the full quantum probability distribution, as given by the solution of the master equation, and which also coincides with the average over all trajectories, does not show the same localization effect. This is not surprising since different trajectories are added incoherently and so one should not expect destructive interference.


\item In Fig.~\ref{fig5} (a) we see that at very short times ($t \lesssim 4$) the semi-classical (TWA) and quantum evolution both agree. The quantum and semi-classical curves then diverge, but the various quantum curves continue to agree with each  other until the time $\tau_D \approx 15 $ irrespective of $\kappa$. This {\it dwell time} is a characteristic of open chaotic systems~\cite{dwell,dwell2} and represents the period during which the open (chaotic) system is transparent to decoherence and evolves similarly to the corresponding closed system. Normally, during the dwell time the system evolves essentially classically and is therefore independent of $\hbar$~\cite{dwell2}. In our case we again find a breakdown from the typical behaviour as the semi-classical broadening diverges from the quantum one long before the dwell time is over. 

\item At times greater than the dwell time, both Fig.~\ref{fig5} (a) and (b) show that the isolated quantum system $(\kappa=0)$ diverges from the open ones ($\kappa \neq 0$). However, the latter show only a very weak dependence on $\kappa$ (provided that it is not vanishingly small). This is particularly true in the case of Fig.~\ref{fig5} (a). A phenomenon that is perhaps related to this effect is known from studies of Loschmidt echoes in time-reversed open quantum systems where it was discovered experimentally \cite{levstein98}, and subsequently understood theoretically \cite{jalabert01}, that there is a universal regime in which the effects of decoherence are dependent only upon the dynamics of the underlying closed classical system and are, paradoxically, independent of the system-environment coupling strength $\kappa$. This regime is known as the {\it Lyapunov regime}, the name indicating that decoherence is dependent only upon the Lyapunov exponents characterizing entropy growth in the underlying classical chaotic system \cite{zurek94}.

\end{enumerate}

\begin{figure}[h]
\centerline{\includegraphics[width=8cm]{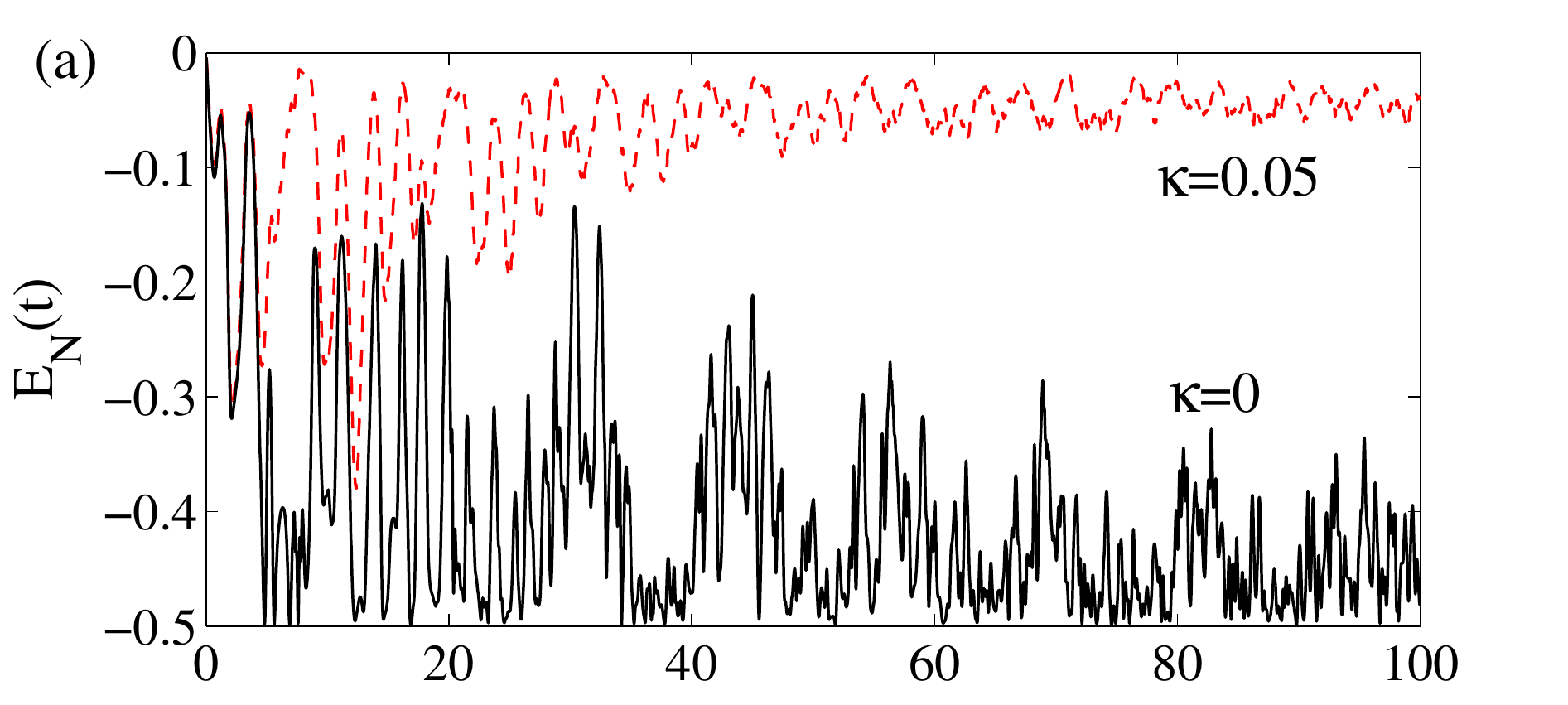}}
\centerline{\includegraphics[width=8cm]{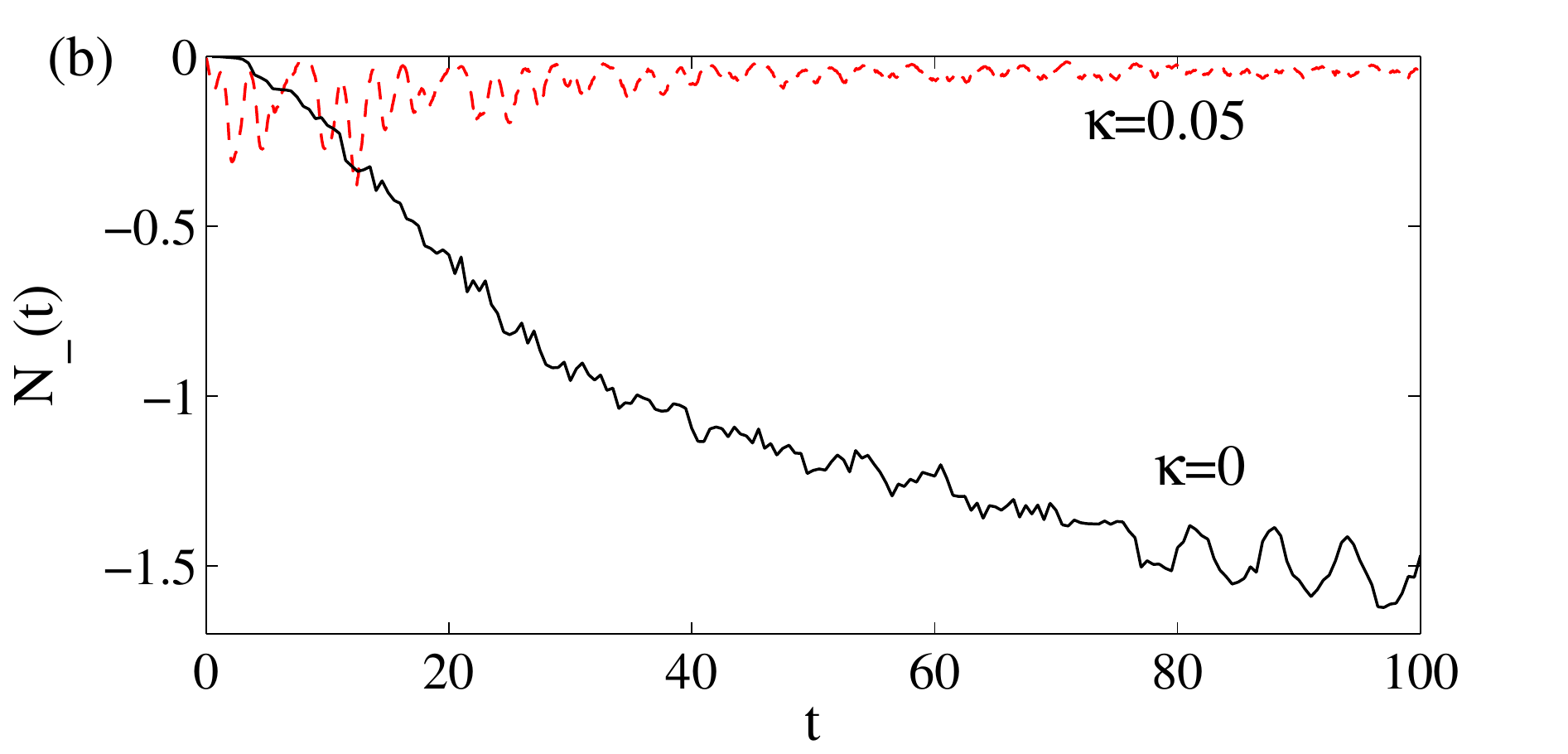}}
\caption{(Color online) The evolution of the negativity $E_N(t)$ (a) and negative fraction $N_-(t)$ (b) for $\kappa=0$ (black solid lines) and $\kappa=0.05$ (red dashed lines). For larger times, the negativity follows a similar behavior with declining oscillation amplitudes, i.e. the bipartite system remains entangled even for long times despite the non-zero $\kappa$ (verified numerically but not shwn here). Decoherence suppresses entanglement, but a substantial fraction survives even large decay losses. Similar behaviour is encountered for the negative fraction of the Wigner distribution as demonstrated in the lower plot. However, it seems that $N_-(t)$ is more sensitive to decoherence. The remaining parameters are as in Fig. \ref{fig1}. } \label{fig6}
\end{figure}

One explanation for the failure of the semi-classical description to reproduce the fully quantum ones even in the presence of decoherence comes from the fact that some quantum properties are maintained despite the decoherence. Since the state $\hat{\rho}(t)$ is in general mixed, standard entanglement measures like the {\it von Neumann entropy} cannot be applied to estimate qubit-field entanglement. Neither can the {\it concurrence} be utilized since the two sub-systems do not constitute two qubits. As a measure of entanglement we use instead the {\it negativity}~\cite{negativ} 
\begin{equation}\label{neg}
E_N(\hat{\rho})=-\sum_i\frac{|\lambda_i|-\lambda_i}{2},
\end{equation}
where $\lambda_i$ is the $i$th eigenvalue of the partial transpose $\hat{\rho}^{T_A}$ of $\hat{\rho}$~\cite{partial} (here ``$A$'' represents one of the two subsystems - qubit or boson mode). The definition (\ref{neg}) means that the negativity is the sum of negative eigenvalues $\lambda_i$. In principle, $E_N(\hat{\rho})<0$ is only a necessary (but not sufficient) condition for entanglement, but, nevertheless, we will take it as an indicator of quantum correlations. We have a lower bound $E_N(\hat{\rho})\geq1/2$. The negativity for a relatively large $\kappa$ and for early times is displayed in Fig.~\ref{fig6} (a). At later times, our numerics indicate that $E_N(\hat{\rho})$ shows rapid variations around the value $\sim-0.05$ and never decays to zero. Thus, the entanglement is {\it not} completely destroyed by the decoherence. It has been shown that for closed quantum systems, non-zero entanglement can be a signature of quantum chaos~\cite{chaudhury09,chaosent}. In general, chaotic dynamics shows large entanglement in closed systems. 

Typically, given a Lindblad term $\mathcal{L}[\hat{A}]$ with $[\hat{A},\hat{H}]=0$ the state $\hat{\rho}(t)$ tends to relax to a steady state which is the ground state of the {\it parent Hamiltonian} $\hat{A}^\dagger\hat{A}$~\cite{diehl}. Here, since $[\hat{x},\hat{H}]\neq0$ there is an interplay between the two terms, unitary Hamiltonian time evolution and non-unitary decoherent time evolution, and the state is not expected to approach some pure steady state. Indeed, the state $\hat{\rho}(t)$ is highly mixed as can be verified by calculating the purity $P(t)=\mathrm{Tr}[\hat{\rho}^2(t)]$~\cite{purity}. Although not shown here, we have calculated $P(t)$ for different $\kappa$'s and only in the special case where $\kappa=0$ is the state pure, i.e.\ $P(t)=1$ for $\kappa=0$ and $P(t)<1$ otherwise.

As mentioned above, another signature of non-classical features is the negative ratio $N_-(t)$ of the Wigner function~\cite{nwig}. The rule-of-thumb is that decoherence tends to suppress non-classical effects, such as; squeezing, sub-Poissonian photon statistics, and coherent superpositions. We also saw an example of this in Fig.~\ref{fig3}. Examples of the negative fraction $N_-(t)$ are presented in Fig.~\ref{fig6} (b). While $N_-(t)$ is suppressed, there is always a fraction that survives. Compared to the negativity, $N_-(t)$ seems somewhat more sensitive to decoherence.

\begin{figure}[t]
\centerline{\includegraphics[width=8cm]{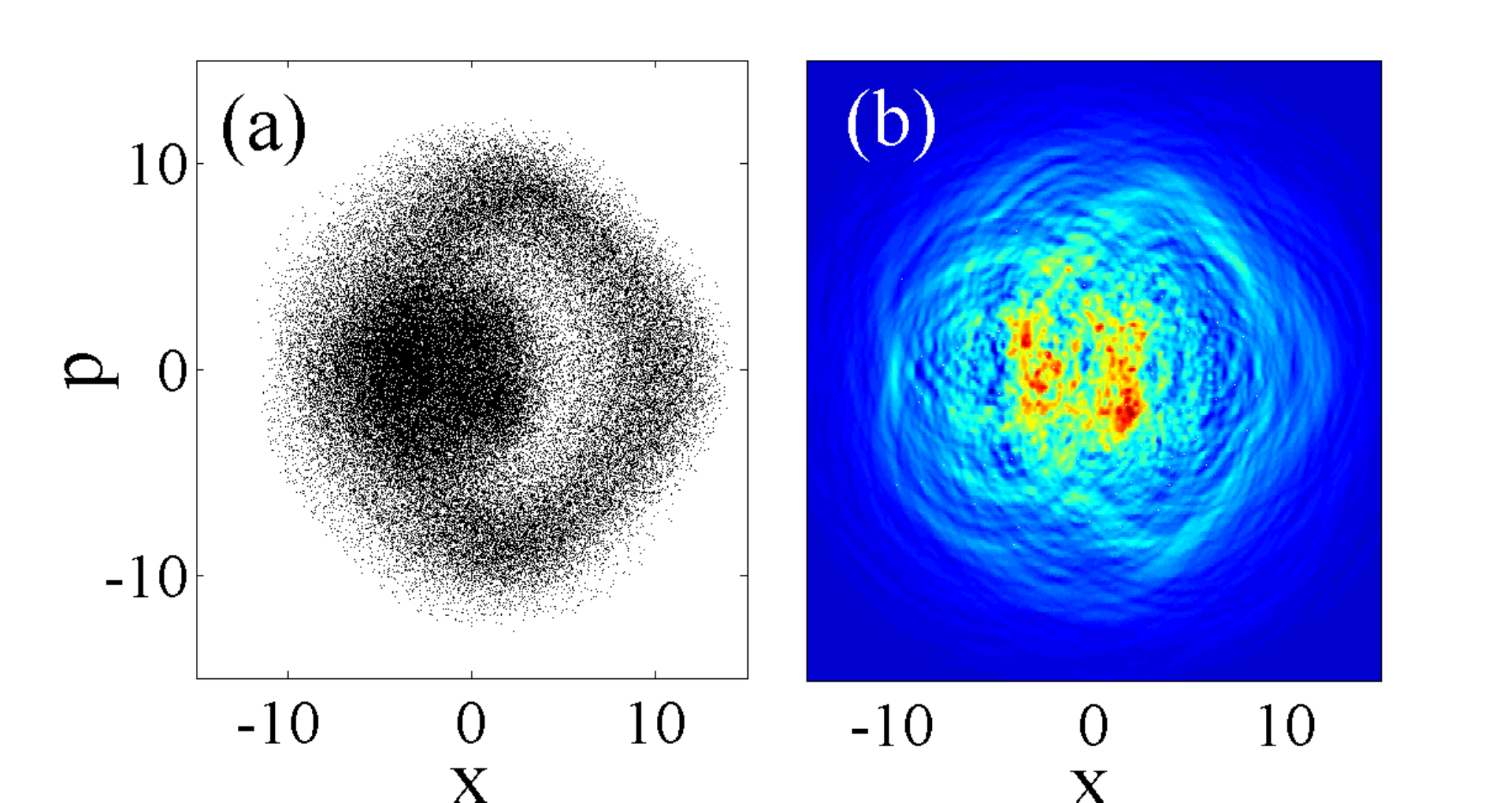}}
\caption{(Color online) Time averaged phase space dynamics: (a) the time-averaged TWA for the resonator field and (b) the corresponding time-averaged Wigner function for $\kappa=0$. In both cases the time average is over the interval $0\leq t\leq 200$. The rest of the dimensionless parameters are as in Fig.~\ref{fig1}. The time-averaged TWA shown here should be distinguished from the stroboscopic maps shown in Fig.~\ref{fig1}. Here data is collected and plotted throughout the evolution time, whereas there only the final data are shown.} \label{time_av}
\end{figure}

We have made much in this paper of the discrepancy between the quantum and classical behaviour of our system. However, it is worth noting the following empirical fact: when the dynamics in phase space is \emph{averaged over time} we find that classical and quantum come back into correspondence. This is illustrated in Fig.~\ref{time_av} where it can be seen that the total region of phase space explored by the TWA during the dynamics is very similar to the total region of phase space explored by the Wigner distribution, even though, as we saw above in Fig.\ \ref{fig5}, at any instant in time their distributions in phase space are quite different. Furthermore, similar looking structures seem to be present in both the quantum and classical distributions. It is noteworthy that for the quantum (Wigner) case in Fig.\ \ref{time_av} we have used unitary time evolution ($\kappa=0$), and so from our previous discussion one might otherwise expect the two distributions to be very different. Instead, the quantum evolution obeys the dynamical limits set in phase space by the classical evolution. In a closed system we might expect this correspondence on the grounds of energy conservation, but here we have an open system which is externally pumped and there is no such argument here. This correspondence between the time-averaged quantum and classical dynamics has been noted before, e.g.\ in experiment \cite{chaudhury09} that reconstructed the time evolution of the Wigner function from measurements. They found that  the time averaged quantum distribution spread throughout phase space but respected the boundaries between the regular and irregular regions with high fidelity.

\subsubsection{Detection}
We have seen examples of how the effects of decoherence come into play for chaotic quantum dynamics. Therefore, a most relevant question concerns experimental observation of both quantum chaotic dynamics itself, and the effect decoherence has upon it. Thus far we have considered the phase space distributions and especially their widths. The full phase space distribution is in principle attainable experimentally via state tomography~\cite{wignermeas,wignermeas0} (see also \cite{chaudhury09}). Such homodyne detection relies on experimental repeatability, i.e.\ repeated measurements on copies of equally prepared systems. Since the system is manifestly chaotic and non-linear, small fluctuations in the preparation will cause large variations in the evolved state and the measured results may seem random. The measurement result will therefore more likely represent a state averaged over many ``realizations''. This, however, is in itself a signature of the chaotic nature. Thus, state tomography should render a phase space distribution with seemingly random structures.   

We may ask whether there are other quantities that are more easily obtained and still show chaotic characteristics? It turns out that even the statistical properties of the intensity of the output cavity field contain information of the quantum chaotic behaviour. The phase space distribution, apart from spreading, will bunch back and forth along the $x$-direction. When passing the center at $x=0$, population transfer between the two qubit components takes place, i.e. adiabaticity breaks down. For short times, when the wave packet is still localized, this will be manifested as sudden jumps in $\langle\hat{\sigma}_z\rangle$. After the delocalization, the clear transitions between the two components turn into rapid fluctuations. This will also be reflected in large fluctuations in the field intensity $\langle\hat{a}^\dagger\hat{a}\rangle$. Classically, the ergodic evolution implies that a phase space trajectory frequently intersects all accessible ``rings'' with radii $r=\sqrt{p^2+x^2}/2$ in phase space. Thus, a classical solution spends some time at all accessible distances from the origin of phase space, and such chaotic motion results in pronounced field intensity fluctuations. For a coherently driven system, the intensity of the output field at time $t$, $\langle\hat{a}_\mathrm{out}^\dagger\hat{a}_\mathrm{out}\rangle_t$, leaking through the cavity mirror is proportional to the cavity mode intensity $\langle\hat{a}^\dagger\hat{a}\rangle_t$~\cite{wm}. Consequently, direct photon detection of the output field gives a handle on the cavity photon fluctuations. 

\begin{figure}[h]
\centerline{\includegraphics[width=8cm]{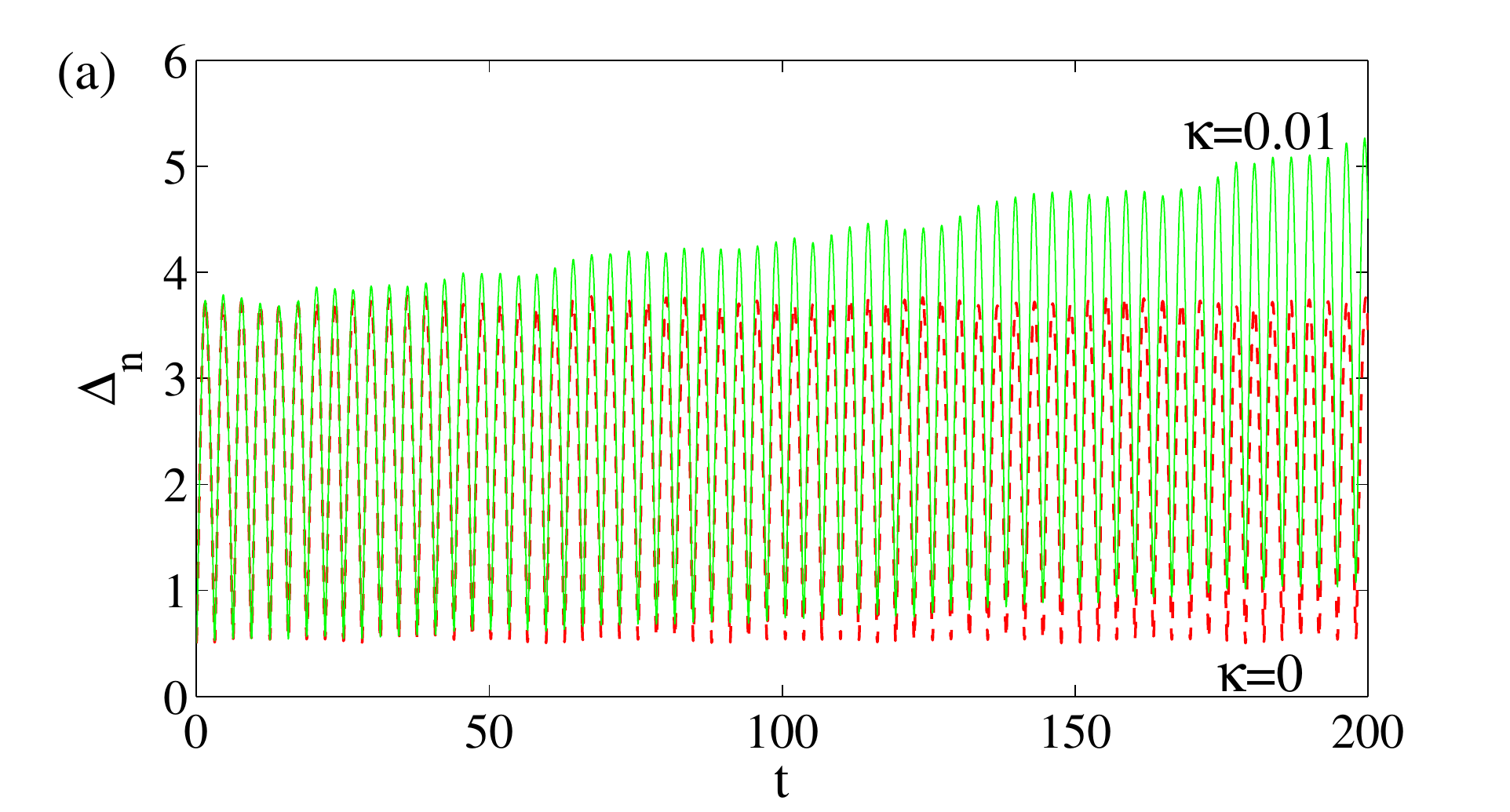}}
\centerline{\includegraphics[width=8cm]{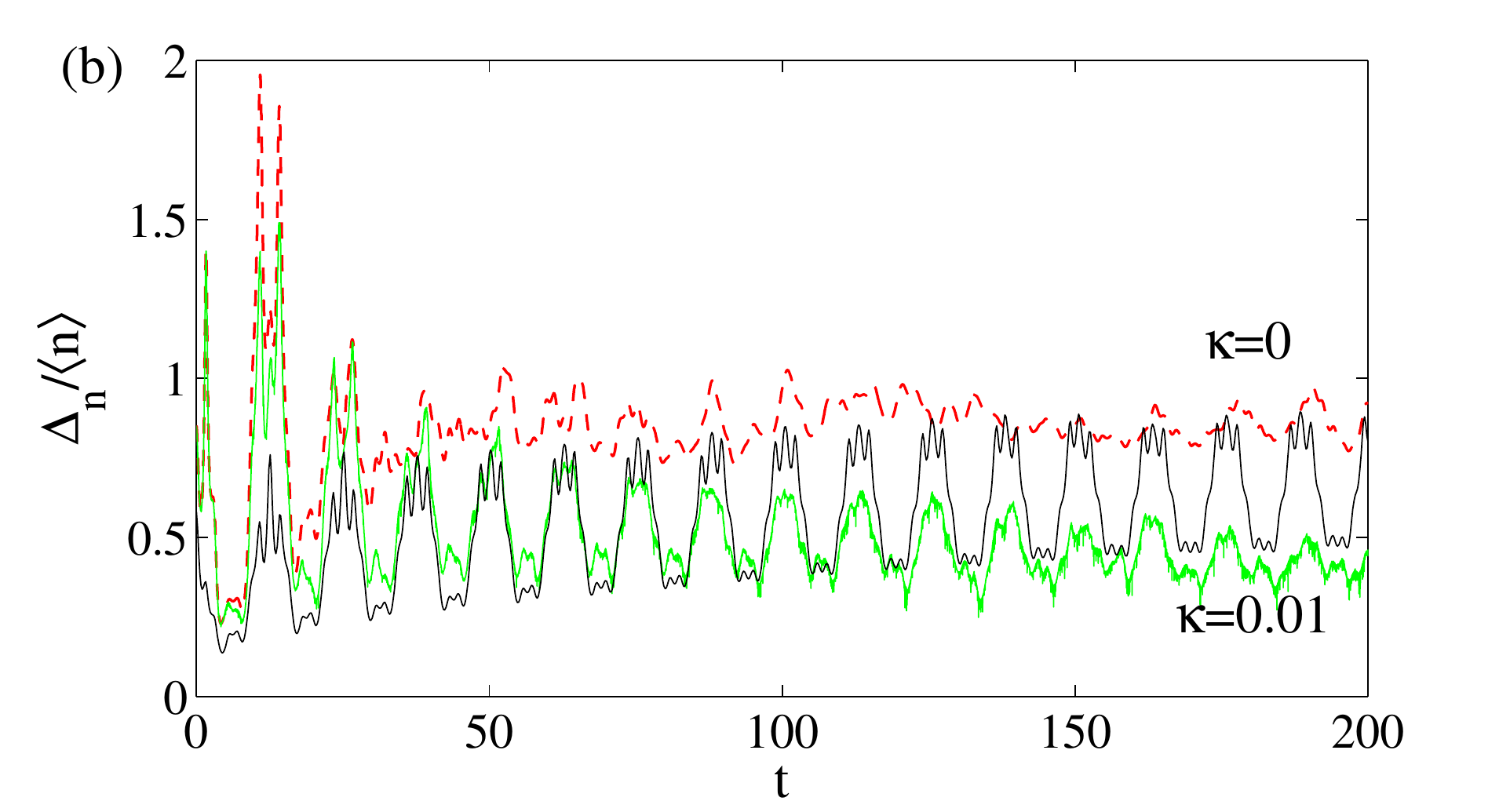}}
\caption{(Color online) Variance of the photon intensity. The upper plot (a) gives the variance for a system pumped with a constant amplitude, i.e. the dynamics is regular. The lower figure (b) shows instead the scaled variance for the system with an oscillating drive. In both plots the red dashed line gives the full quantum result with no decoherence, and the green solid line gives the same except that $\kappa=0.01$. The black solid line displays the corresponding TWA result. Comparing (a) and (b) clearly indicates a more irregular structure in the photon variance whenever the system is evolving in the chaotic regime. The rest of the parameters are as in Fig.~\ref{fig1}.} \label{fig7}
\end{figure}

In Fig.~\ref{fig7} we compare the photon variance of the full field state $\hat{\rho}_f(t)$ (i.e.\ not for single trajectories) for a constantly pumped system (a), with the one driven with an oscillating amplitude (b). In (a) we show the variance $\Delta_n=\sqrt{\langle\left(\hat{a}^\dagger\hat{a}\right)^2\rangle-\langle\hat{a}^\dagger\hat{a}\rangle^2}$ and in (b) the scaled variance $\Delta_n/\langle \hat{a}^\dagger\hat{a}\rangle$. The reason why we do not divide the variance in the regular case (a) is because $\langle\hat{a}^\dagger\hat{a}\rangle$ periodically becomes approximately zero and would cause divergences. For the constant driving case we find a nicely oscillating behaviour, while in the chaotic case the variance shows much greater fluctuations. Another difference is that decoherence enhances field intensity fluctuations in the regular regime, while it suppresses it in the chaotic regime. As a comparison we also display the TWA result. As is seen, without decoherence the quantum variance exceeds the semi-classical one. For strong enough decoherence, the quantum variance may actually be smaller than the semi-classical variance.

\section{Conclusion}\label{sec5}
In this paper we have proposed using a circuit QED setup for studies of quantum chaotic dynamics in a system with no clear classical limit. Special attention was paid to the influence of decoherence deriving from continuous quadrature measurements of the resonator field. Since the measurement back-action only induces decoherence and no dissipation, the corresponding semi-classical equations of motion are transparent to this effect. We demonstrated that the set of semi-classical equations of motion predict classical chaos. In general, decoherence, as it is known to diminish quantum coherence, may explain how classical chaos can appear from quantum mechanics. This is typically the case for systems with a classical limit. However, we showed that this method of achieving classical-quantum correspondence fails in our bipartite model where one of the two subsystems is manifestly quantum. In particular, within a wide range of decoherence rates $\kappa$ we found that the semi-classical results could not be retrieved (except by averaging over time). By calculating the negativity we concluded that this discrepancy between quantum and semi-classical chaotic evolution could originate from the presence of truly quantum properties such as entanglement even in the presence of strong decoherence. While the full quantum state $\hat{\rho}(t)$ did not show much localization, the states appearing from `selective measurements' gave a form of localization distinct in origin from standard dynamical localization but actually rather close in mechanism to the stochastic Schr\"{o}dinger equation we employed to describe the measurements. We further argued that the chaotic nature of the system should show up in the photon statistics which are readily measured in experiments. We note that the amplitude of the information transfer rate $\kappa$ should be controllable by adjusting the homodyne detection. 

It should be pointed out that our results apply to other chaotic models exposed to decoherence. Maybe the most relevant example is the Dicke model~\cite{dicke} describing a set of $N$ two-level qubits coherently interacting with a single cavity mode. The ultrastrong coupling regime was recently achieved experimentally in an effective Dicke model~\cite{esslinger}, and by exciting the system sufficiently it can be expected to enter a chaotic regime, where our predictions should be detectable in the output cavity field. Our analysis assumes pure decoherence and no dissipation, but we have also tried other forms of the Lindblad terms including, for example, photon decay, and found similar results. When the system is dissipative, one must keep in mind that photon decay itself may cause a localization of the phase space distribution due to energetic considerations that are different from the localization discussed here.

\begin{acknowledgements}
We thank Alexander Altland, Shohini Ghose and G\"oran Lindblad for enlightening discussions, and anonymous referees for useful comments and suggestions. JL acknowledges support from the Swedish research council (VR), Kungl. Vetenskapsakademien (KVA), and Deutscher Akademischer Austausch Dienst (DAAD). DO acknowledges support from NSERC (Canada).
\end{acknowledgements}

\appendix

\section{The individual distributions in the position and momentum quadratures}\label{app:residuals}

\begin{figure}[h]
\centerline{\includegraphics[width=8cm]{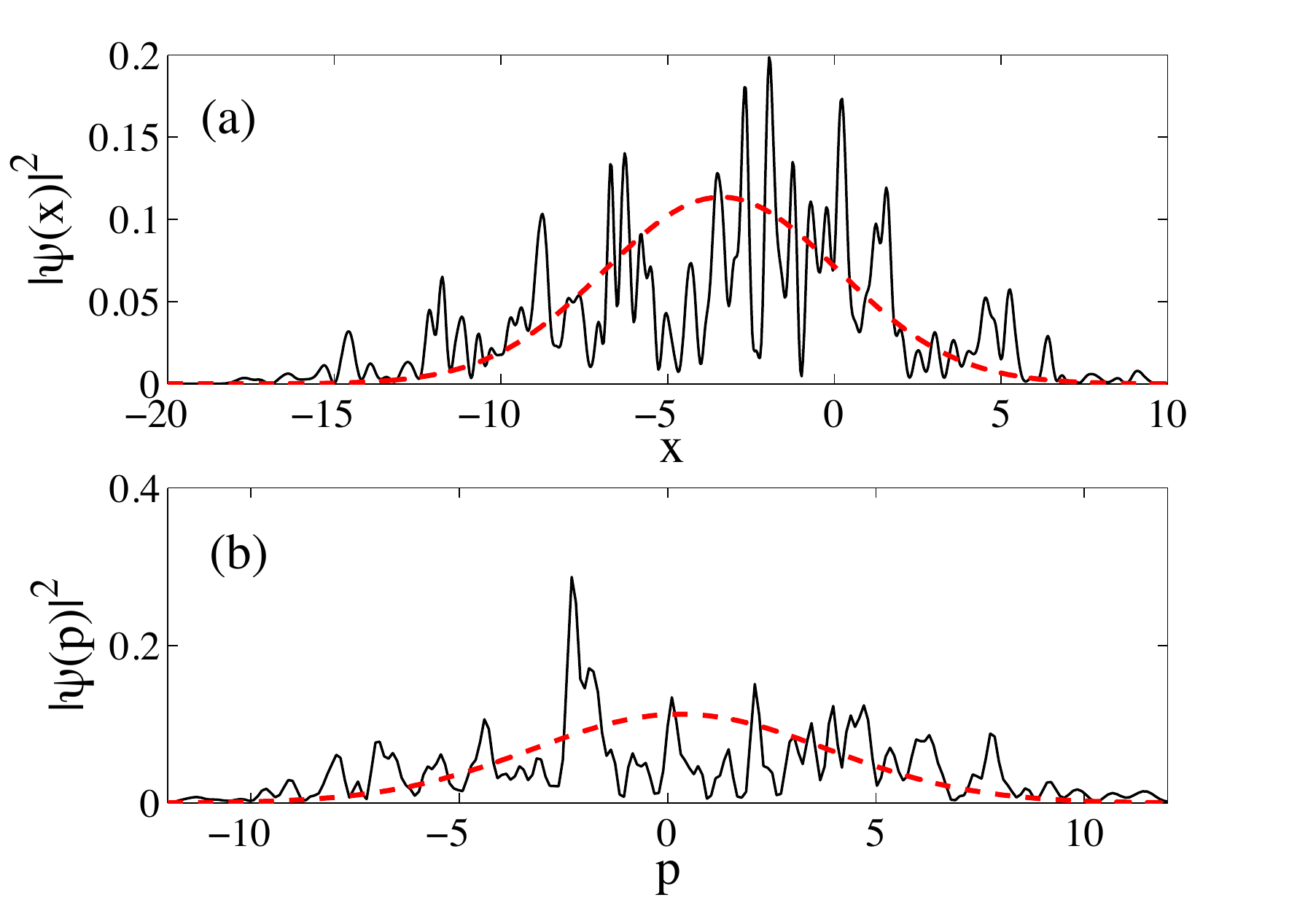}}
\caption{(Color online) The $x$ and $p$ probability distributions from the fully quantum evolution for $\kappa=0$.  (a) the $x$-quadrature and (b) the $p$-quadrature of the field. The black solid lines give the actual data and the red dashed lines are fits to gaussians with the same means and standard deviations as the data. The time is here $t=200$ and the remaining parameters are as in Fig. \ref{fig1}. } \label{quadraturedistributionsk0}
\end{figure}

\begin{figure}[h]
\centerline{\includegraphics[width=8cm]{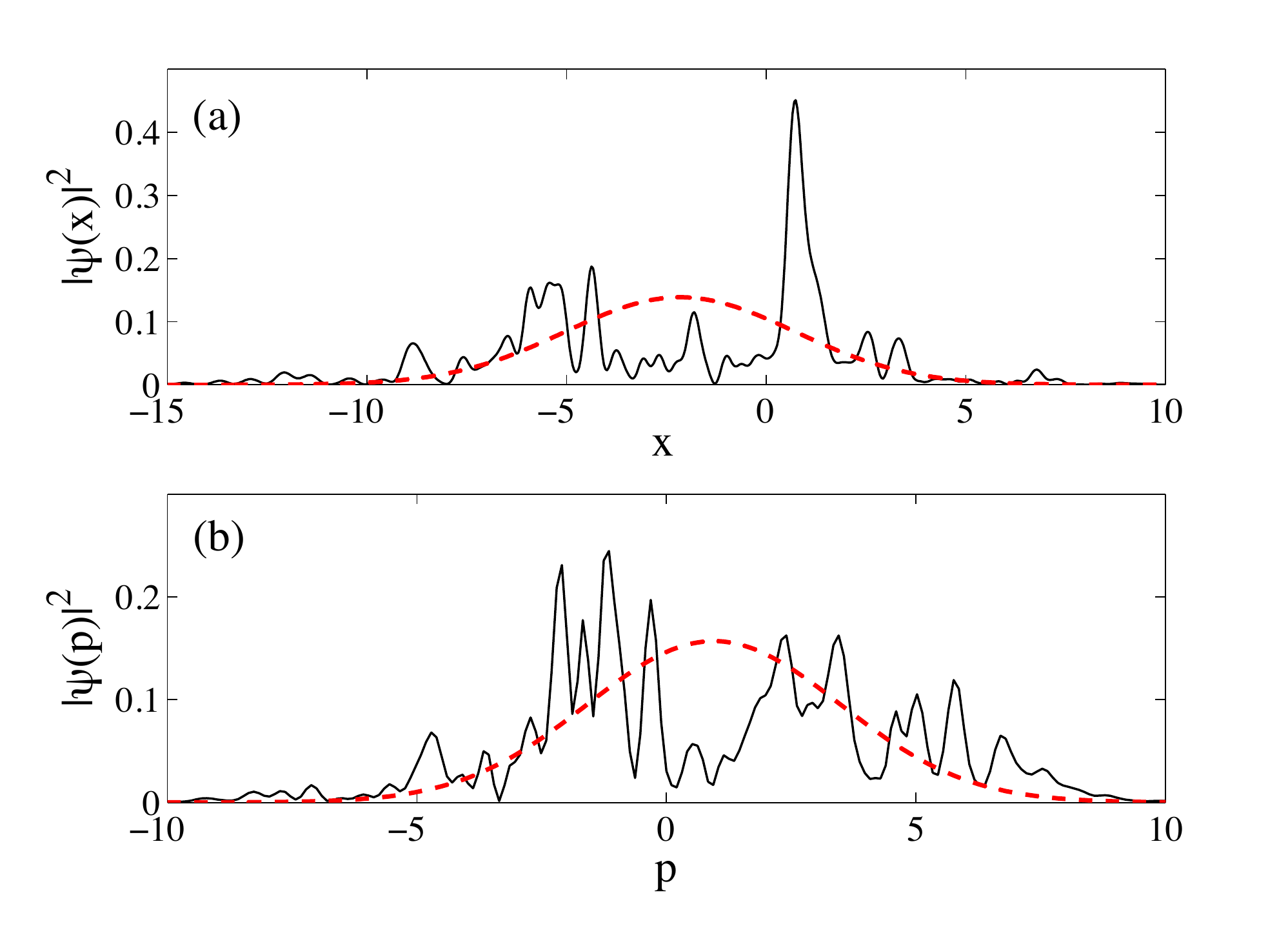}}
\caption{(Color online) Same as Fig.~\ref{quadraturedistributionsk0} but for $\kappa=0.05$.} \label{quadraturedistributionsk05}
\end{figure}

\begin{figure}[h]
\centerline{\includegraphics[width=8cm]{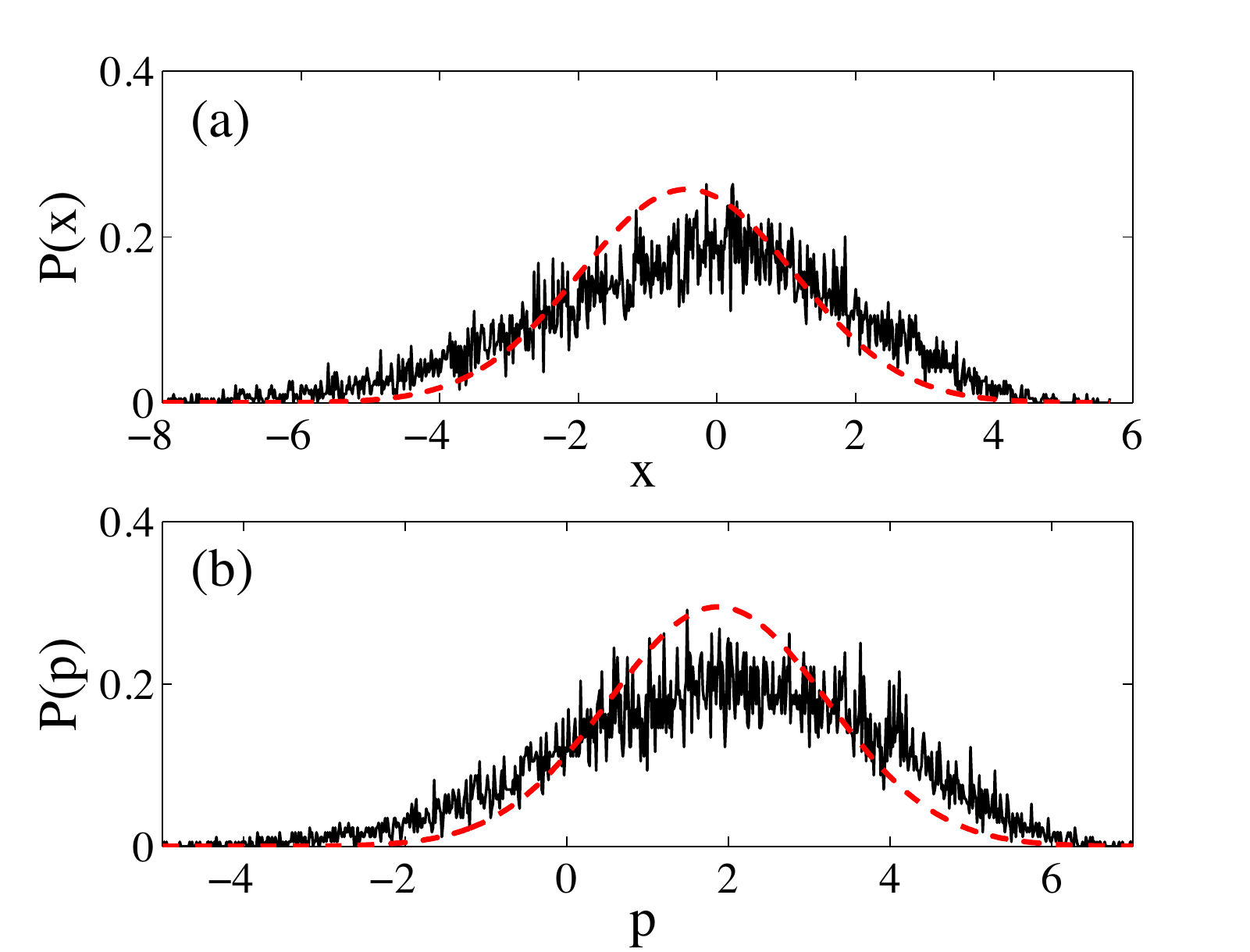}}
\caption{(Color online) Same as Fig.~\ref{quadraturedistributionsk0} but for the TWA evolution. The rapid variations are typical for semi-classical distributions were quantum pressure does not forbid sub-Planck structures.} \label{quadraturedistributionsTWA}
\end{figure}

In Section \ref{sec4} in the main text we presented results concerning the area $\Delta_{xp}$ in phase space  occupied by the quantum and classical distributions (and also the area $\delta_{xp}$ associated with a sum over single stochastic trajectories). However, since our measurement scheme only measures the $x$-quadrature it is interesting to know what the typical behaviour of each individual quadrature (the so-called residuals of the phase space probability distribution) is rather than just the combined uncertainty shown in Fig.\ \ref{fig5}. For this reason we display in Figs.\ \ref{quadraturedistributionsk0},  \ref{quadraturedistributionsk05}, and \ref{quadraturedistributionsTWA} snapshots of the the probability distributions for the two field quadratures individually at time $t=200$. It is interesting to note that there is more fine structure present in the quantum probability distribution for $\kappa=0$ than for $\kappa=0.05$, which is a result of the suppression of quantum interference. In the semi-classical case, the smoothening effect due to the the quantum diffusion $\mathcal{L}_\mathrm{qdiff}$ is absent resulting in much finer structures, see Fig.~\ref{quadraturedistributionsTWA}.

\end{document}